\documentclass[twocolumn,aps,showpacs,prb,amsmath,amssymb,floatfix,superscriptaddress]{revtex4}
\usepackage{color}
\usepackage{graphicx}
\usepackage{dcolumn}
\usepackage{bm}
\usepackage{array}
\usepackage{float}
\usepackage{supertabular}
\usepackage{longtable}
\begin{document}

\newcommand{\lprime}{l\hspace{-.3mm}'}

\title{General method for calculating the universal conductance of strongly correlated junctions of multiple quantum wires}

\author{Armin Rahmani}
\affiliation{Department of Physics, Boston University, Boston, MA 02215 USA }
\author{Chang-Yu Hou}
\affiliation{Instituut-Lorentz, Universiteit Leiden, P.O. Box 9506, 2300 RA Leiden, The Netherlands}
\author{Adrian Feiguin}
\affiliation{Department of Physics and Astronomy, University of Wyoming, Laramie, Wyoming 82071, USA}
\author{Masaki Oshikawa}
\affiliation{Institute for Solid State Physics, University of Tokyo, Kashiwa 277-8581, Japan}
\author{Claudio Chamon}
\affiliation{Department of Physics, Boston University, Boston, MA 02215 USA }
\author{Ian Affleck}
\affiliation{Department of Physics and Astronomy, University of British Columbia, Vancouver, B.C., Canada, V6T 1Z1}

\date{\today}

\begin{abstract}
We develop a method to extract the universal conductance of junctions of multiple quantum wires, a property of systems connected to reservoirs, from static ground-state computations in closed finite systems. The method is based on a key relationship, derived within the framework of boundary conformal field theory, between the conductance tensor and certain ground state correlation functions. Our results provide a systematic way of studying quantum transport in the presence of strong electron-electron interactions using efficient numerical techniques such as the standard time-independent density-matrix renormalization-group method. We give a step-by-step recipe for applying the method and present several tests and benchmarks. As an application of the method, we calculate the conductance of the M fixed point of a Y junction of Luttinger liquids for several values of the Luttinger parameter $g$ and conjecture its functional dependence on $g$.

\end{abstract}

\maketitle

\section{Introduction}
\label{section:Introduction}

Advances in molecular electronics can extend the limits of device
miniaturization to the atomic scales where entire electronic circuits
are made with molecular building blocks.~\cite{Nitzan03,Tao06} Single
molecule junctions connected to two macroscopic metallic leads have
already been successfully fabricated,~\cite{Reed97} and there are several proposals such as laying quantum wires on top of each other for making junctions of multiple quantum wires.~\cite{Kushmerick02}

\begin{figure}[htb]
\centering
\includegraphics[width = 8 cm]{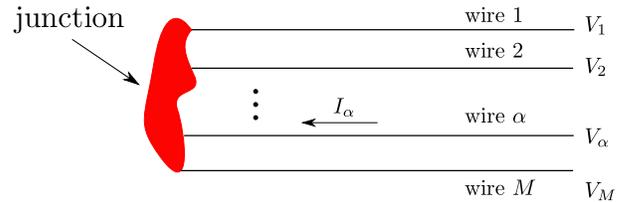}
\caption
{
(Color online) A schematic illustration of the generic system that is the subject of this paper. We have $M$ quantum wires connected to a molecular system of certain structure and interactions. We have currents $I_\alpha$ running in wires $\alpha=1\dots M$ and voltages $V_\alpha$ applied to the endpoints of the wires.}
\label{fig:schematic}
\end{figure}

If we eventually manage to build entire electronic circuits with molecular building blocks, a paramount goal in the field of molecular electronics, junctions of three or more quantum wires will inevitably be a key ingredient. These junctions are comprised of several quantum wires, i.e., quasi-one-dimensional (1D) metallic structures with atomic scale sizes, that are connected to one another by a given molecular structure as shown schematically in Fig.~\ref{fig:schematic}. The structure and interactions at the junction depend on the particular system under study. What we mean by metallic in the above description of quantum wires is that they are capable of conducting electricity due to the presence of gapless excitations. A generic description for these 1D quantum wires is based on the Tomonaga-Luttinger-liquid theory.~\cite{Tomonaga50,Luttinger63,Mattis65,Haldane81} Structures involving Luttinger-liquid quantum wires have been the subject of numerous recent studies.~\cite{Nayak99,Lal02,Rao04,Chen02,Egger03,Pham03,Safi01,Moore02,Yi98,Kim04,Furusaki05,Giuliano05,Kazymyrenko05,Das08,Agarwal09,Bellazzini09a,Bellazzini09b,Safi09a, Safi11,Aristov10,Aristov11} Experimentally, such quantum wires are realized with carbon nanotubes or through the cleaved edge overgrowth technique in GaAs heterostructures.~\cite{Bockrath99,Yao99,Steinberg08} Electrical current running in the wires can pass through this molecular structure at the junction. 

At molecular length scales, quantum mechanics is important and the system represented above must be modeled accordingly. The simplest theoretical description of systems such as this is based on the tight-binding model, with anti-commuting creation and annihilation operators introduced for different atomic sites. An effective Hamiltonian can then be written in terms of these creation and annihilation operators and generically involves hopping terms $c^\dagger c$ and density-density interaction terms $nn$ with $n=c^\dagger c$.

Suppose we wish to study a rather arbitrary junction, the structure and interactions of which are represented by a tight-binding Hamiltonian. A very basic question regarding this system is how it conducts electricity. Consider a system of $M$ wires. In the presence of voltage biases $V_\alpha$ applied to the endpoint of wire $\alpha$ for $\alpha=1 \dots M$, current $I_\alpha$ will flow along each wire $\alpha$. By convention, a current $I_\alpha$ is positive if it flows toward the junction and negative if it flows away from it as seen in Fig.~\ref{fig:schematic}. In general, i.e., at arbitrary biases and temperatures, this problem is very complicated and the currents flowing in the quantum wires are nonuniversal functions of the temperature, the voltages $V_\alpha$, and the microscopic details of the system. In this paper, we are concerned with the linear-response regime where universal behavior can emerge. We work at zero temperature and consider only the limit of infinitesimal biases. The currents in this regime will be a linear combination of the applied biases as seen in the following:
\begin{equation}\label {eq:G_definition}
 I_\alpha =\sum_\beta G_{\alpha \beta} V_\beta.
\end{equation}
This linear relationship defines the linear conductance tensor $G_{\alpha \beta}$ which is the quantity of interest in this paper.

One of the most important ideas of modern physics is the remarkable universality that emerges near critical points. It turns out that due to the criticality of the bulk of quantum wires, i.e., having divergent correlation lengths, a large degree of universality also emerges in the behavior of quantum junctions. The universality can be understood in the framework of the renormalization group (RG).~\cite{Wilson75} One can argue that in the limit of small biases and low temperatures, many of the microscopic details of the junction are irrelevant in the RG sense, which means their contribution to conductance, and other physical observables, decays to zero at large distances and low energies. 

The junctions we are concerned with in this paper fall into the category of quantum impurity problems. The junction, with all the complex structure and interactions it contains, is localized at the endpoints of the wires. It can therefore be thought of as one (rather arbitrary) impurity inserted into a system, the bulk behavior of which is given by that of $M$ independent quantum wires. A classic example of quantum impurity problems is the Kondo model describing the behavior of conduction electrons interacting with a local magnetic moment.~\cite{Kondo64} The powerful methods of boundary conformal field theory (BCFT) have proven useful in a multitude of quantum impurity problems.~\cite{Affleck90,Affleck10} Thus, BCFT is the main analytical technique used in this paper.

Determining the conductance of quantum junctions in the presence of strong electron-electron interactions is a long-sought and challenging goal. The Landauer-B\" uttiker's formalism, which is the method of choice in the calculation of quantum conductance, does not account for these
interactions, which indeed play a key role in low dimensions. Functional renormalization-group methods have been helpful in studying the interaction effects in the vicinity of the junction, but their applicability is also dependent upon the presence of large noninteracting leads.~\cite{Barnabe05a,Barnabe05b}

In recent years, efficient numerical methods, such as the density-matrix renormalization group (DMRG),~\cite{White92} have been developed for studying strongly correlated quasi-1D quantum problems. Since the quantum junctions described above can be thought of as quasi-1D (by folding all the wires to one side so they run parallel to one another), these numerical methods could potentially be efficient tools for computing the conductance of junctions with an arbitrary number of wires and in the presence of strong interactions.

In fact, DMRG has already been applied to the study of quantum junctions.~\cite{Guo06} However, when it comes to calculating the conductance of strongly correlated junctions, there are fundamental difficulties even when we are armed with powerful tools such as DMRG. One such difficulty arises from the fact that conductance is a property of an open quantum system. We define the conductance in terms of the current passing through the system and the underlying assumption is that we have reservoirs that can act as sources and drains for electrons. To study conductance, we either need to model the reservoirs carefully or send them to infinity. The latter is a simpler and more elegant way of formally dealing with quantum transport, but has the downside that for a numerical calculation of conductance, we would need to model large enough systems that faithfully approximate the semi-infinite ones.

Another difficulty with calculating the linear conductance is that, within the linear-response framework, conductance is formally related to dynamical correlation functions. It may then appear that one needs to use the much more computationally demanding time-dependent numerical methods such as time-dependent DMRG to calculate the conductance. 

For junctions of two quantum wires, time-dependent DMRG has already been used for conductance calculations.~\cite{Cazalilla02,Luo03,White04} A brute force calculation with time-dependent methods in large systems is not, however, currently feasible for strongly correlated junctions of more than two quantum wires.

It is the objective of this paper to make such calculations possible with a combination of analytical and numerical techniques. More specifically, the main objective of this paper is to develop a formalism that would allow us to apply numerical methods such as time-independent DMRG and the related matrix product states to calculate the linear-response conductance of strongly-correlated junctions of an arbitrary number of quantum wires with rather generic structures and interactions in the junction.~\cite{Verstraete04} In this paper, we focus on the systems with spinless electrons, but our method can also be extended to systems with spin-1/2 electrons. 

One particular application for the formalism we seek to develop in this paper is the problem of the M fixed point in a Y junction of spinless Luttinger liquids. The existence of this nontrivial fixed point was conjectured many years ago, but its nature, and more specifically its conductance, had remained an open question in quantum impurity problems.~\cite{Chamon03,Oshikawa06}

The outline of this paper is as follows. In Sec.~\ref{sec:Main_Results}, we summarize the main results of this paper. We present a key relationship [Eq.~(\ref{eq:key_relation_intro})] between the junction conductance and certain static correlation functions in a finite system. This relationship serves as the basis of the method developed here. We also present in Sec.~\ref{sec:Main_Results} a step-by-step recipe for applying the method in practice as well as a summary of the new results on the Y junction, which were obtained with this method. In Sec. \ref{section:Noninteracting_Conductance}, we present some explicit calculations in a noninteracting lattice model that help motivate the derivation of Eq.~(\ref{eq:key_relation_intro}) and clarify the connection of the continuum results to lattice calculations. The results derived are in fact some special cases, which can be obtained with elementary methods, of the general relation Eq.~(\ref{eq:key_relation_intro}), the derivation of which requires the machinery of BCFT. In Sec.~\ref{section:Bosonization_CFT}, we briefly review the main analytical techniques, namely, bosonization and boundary conformal field theory, used in this paper and set up the notation. Section~\ref{section:Conductance_Correlations} is devoted to deriving Eq.~(\ref{eq:key_relation_intro}) in the BCFT framework. In Sec.~\ref{section:Method}, we discuss in detail the method proposed in this work for conductance calculations and clarify practical issues regarding a lattice-model implementation. In Sec.~\ref{section:Benchmarks}, we present numerical benchmarks with DMRG for interacting systems and exact diagonalization for noninteracting systems to verify the correctness of the method. In Sec.~\ref{section:Y_Junction}, we study a Y junction of quantum wires and obtain the previously unknown conductance of its M fixed point as a function of the Luttinger parameter $g$. Finally, we conclude in Sec.~\ref{section:conclusions} by outlining the outlook for future applications and the impact of the results obtained in this paper. Some of the results of this paper have been briefly reported in Ref. 42.

\section {Main results}\label{sec:Main_Results}

We developed a method to extract the universal linear conductance $G_{\alpha \beta}$ of quantum multi-wire junctions defined in Eq.~(\ref{eq:G_definition}) from a calculation of the ground state expectation values of operators involving currents and densities in an appropriately constructed finite system.

\begin{figure}[htb]
\centering
\includegraphics[width = 8 cm]{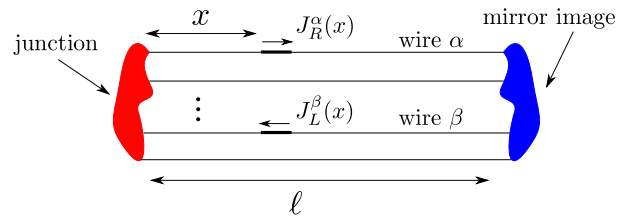}
\caption
{
(Color online) A schematic illustration of the finite system, constructed from the junction of interest and an appropriate mirror image, which we use in our method to extract the conductance $G_{\alpha \beta}$. The conductance is related to the ground state $\langle J_R^\alpha(x)J_L^\beta(x) \rangle_{\rm GS}$ correlation function of chiral currents in this finite system through Eq.~(\ref{eq:key_relation_intro}).}
\label{fig:main_schematic}
\end{figure}

At the core of the method lies an important general relationship, which we recently derived in Ref.~\onlinecite{Rahmani10} using the machinery of BCFT. The relationship is derived in Sec.~\ref{section:Conductance_Correlations} and simply states that
\begin{equation}\label {eq:key_relation_intro}
\lim_{ x \rightarrow\infty}\langle J_R^\alpha(x)J_L^\beta(x) \rangle_{\rm GS}\left[4\: \ell\:
\sin\left( \frac{\pi}{\ell}x\right) 
\right]^{2} \frac{e^2}{h}=G_{\alpha \beta} ,
\end{equation}
where $x$ is the distance from the boundary on the left in a system of length $\ell$, with $\ell \rightarrow\infty$ and finite $x/\ell$, constructed from the junction of interest and an appropriate mirror image placed on the right endpoints of the wires as seen in Fig.~\ref{fig:main_schematic}. Here $J_R^\alpha(x)$ ($J_L^\beta(x)$) is the right-moving (left-moving) current on wire $\alpha$ ($\beta$). Note that although Eq.~(\ref{eq:key_relation_intro}) holds asymptotically ($x \rightarrow\infty$), it can be used to extract the conductance $G_{\alpha \beta}$ even with finite but large enough $x$.

The relationship above is the key ingredient of our method for calculating the conductance. If we can compute the quantity $\langle J_R^\alpha(x)J_L^\beta(x) \rangle_{\rm GS}$, which is a ground-state expectation value in a finite system of length $\ell$, as a function of $x$, then we can multiply it by a universal function to get the left-hand side of Eq.~(\ref{eq:key_relation_intro}) above. This quantity will then saturate to the universal conductance of the junction for large $x$.
  
Apart from the derivation of the key Eq.~(\ref{eq:key_relation_intro}), we provide a recipe for applying this continuum result to a lattice calculation. This requires specifying the procedure for constructing the lattice (tight-binding) Hamiltonian of the aforementioned finite system from the Hamiltonian of the junction of interest (that couples infinitely long wires). It also requires specifying lattice operators, i.e., in terms of the tight-binding fermionic creation and annihilation operators, the correlation functions of which are a good approximation to $\langle J_R^\alpha(x)J_L^\beta(x) \rangle_{\rm GS}$. This is important because the chiral current operators are defined for the continuum theory, and chiral creation and annihilation operators can not be directly modeled on the lattice. The recipe for applying the key relation Eq.~(\ref{eq:key_relation_intro}) to a lattice computation is given in Sec.~\ref{section:Method}.

For quick reference and an illustration of the method, here we give a simple example and explain a step-by-step application of the method to the well-known problem of a weak link in the Luttinger liquid. The starting point for applying our method is always a tight-binding lattice Hamiltonian of spinless electrons for bulk wires and their connection at the junction, namely, $H=H_{\rm boundary}+H_{\rm bulk}$. For a weak link in a Luttinger liquid, we can write 
\[H_{\rm boundary}= -tc^\dagger_{1,0}c_{2,0}-tc^\dagger_{2,0}c_{1,0}\]
and the following Hamiltonian for the bulk of the wires:
\begin{eqnarray}
H_{\rm bulk}=\sum_{\alpha=1}^2\sum_{j=0}^{\infty}\big[-c^\dagger_{\alpha,j}c_{\alpha,j+1}-c^\dagger_{\alpha,j+1}c_{\alpha,j}\\
+V(n_{\alpha,j}-\frac{1}{2})(n_{\alpha,j+1}-\frac{1}{2})\big].\nonumber
\end{eqnarray}
Note that there is some arbitrariness in dividing the system into the junction and wires. The boundary Hamiltonian above is a minimal choice, but including more sites in $H_{\rm boundary}$ does not affect the results as long as the system is large enough and the correlation functions discussed below are computed far away from the boundary.
Given the system Hamiltonian, our method consists of the following steps.
\begin{enumerate}
\item Construct the finite system as in Fig.~\ref{fig:main_schematic}. For this we need to construct a Hamiltonian $H^\prime=H_0+H^\prime_{\rm bulk}+H_\ell$ where $H_0$ and $H_\ell$ respectively describe the junction on the left side of the system ($x=0$) in Fig.~\ref{fig:main_schematic} and the mirror image at $x=\ell$. The recipe for constructing these Hamiltonians is simple. The left boundary Hamiltonian $H_0$ is simply equal to $H_{\rm boundary}$ and the bulk Hamiltonian $H^\prime_{\rm bulk}$ has exactly the same form as the bulk Hamiltonian of the semi-infinite system but a finite number of terms, i.e., $\sum_{j=0}^{\infty}\rightarrow\sum_{j=0}^{N-1}$. The construction of the right Hamiltonian goes as follows. First we consider the same Hamiltonian as $H_{\rm boundary}$ but acting on the other endpoint, namely $ -tc^\dagger_{1,N}c_{2,N}-tc^\dagger_{2,N}c_{1,N}$ and then we apply two transformations $K$ and $C$ on this Hamiltonian. $K$ simply takes the complex conjugate and $C$ changes $c\rightarrow c^\dagger$. In this case, assuming a real hopping amplitude $t$, we have 
\begin{eqnarray*}
H_\ell&=&C(-tc^\dagger_{1,N}c_{2,N}-tc^\dagger_{2,N}c_{1,N})\\
&=&tc^\dagger_{1,N}c_{2,N}+tc^\dagger_{2,N}c_{1,N}.
\end{eqnarray*} 
\item Having constructed the Hamiltonian $H=H_0+H^\prime_{\rm bulk}+H_\ell$ of the finite system, measure, by a numerical DMRG calculation, the following ground state expectation value:
\[
\langle J_R^1(m)J_L^2(m)\rangle=-\frac{1}{2v^2}\langle J^1(m)J^2(m)\rangle,
\]
where $J^\alpha(m)=i\left(c^\dagger_{\alpha,m}\: c_{\alpha,m+1}-c^\dagger_{\alpha,m+1}\: c_{\alpha,m}\right)$ is simply the current operator and $v$ is the charge carrier velocity for the Luttinger liquids described by $H_{\rm bulk}$. The above equation is valid for time-reversal symmetric systems like the example at hand. The general construction of the operator $\langle J_R^1(m)J_L^2(m)\rangle$ in terms of the lattice creation and annihilation operators is given in section~\ref{section:Method}.
\item Fit the data for $\langle J_R^1(m)J_L^2(m)\rangle$ to the asymptotic functional form from Eq.~(\ref{eq:key_relation_intro}), i.e., $\langle J_R^1(m)J_L^2(m)\rangle \propto \left[4\: N\:
\sin\left( {\pi \over N}m\right) 
\right]^{-2}$, and obtain $G_{12}$ from the overall coefficient. 
\end{enumerate}
Note that if $G_{12}=0$, the fitting is tricky. These details will be discussed later.

\begin{figure}[htb]
\centering
\includegraphics[width = 8 cm]{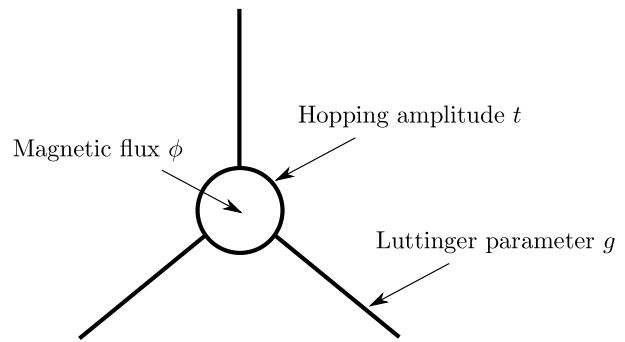}
\caption
{
A schematic illustration of the Y junction numerically studied in section~\ref{section:Y_Junction}. The M fixed point corresponds to the time reversal symmetric case $\phi=0,\pi$. 
}
\label{fig:Yjunction}
\end{figure}

The other important results of this paper concern a concrete application of the method to a previously unsolved quantum impurity problem. These results are presented in Sec.~\ref{section:Y_Junction}. The main problem solved in that section by an explicit numerical DMRG calculation is determining the conductance of a nontrivial fixed point in an interacting Y junction of three quantum wires depicted in Fig.~\ref{fig:Yjunction}. The existence of this fixed point known as the M fixed point was conjectured in Ref.~\onlinecite{Chamon03,Oshikawa06}, but the properties of the fixed point including its conductance remained unknown. In this paper, by combining the method developed in Secs. \ref{section:Conductance_Correlations} and \ref{section:Method} and numerical computations with time-independent DMRG, we calculate the universal conductance of this fixed point as a function of the Luttinger parameter $g$ (a parameter that quantifies the strength of electron-electron interactions in the wires). Based on the numerical results, we claim that the conductance of the M fixed point depends on the Luttinger parameter as
\begin{equation}
G_{12}(g)=G_{21}(g)=-\frac{2 g \gamma}{2g+3\gamma-3g\gamma}{e^2 \over h},
\end{equation}
where $\gamma\approx 0.42$. The above expression is a universal result which, for any nonvanishing hopping amplitude $t$, holds independently of the value of $t$. 
  We also explicitly verify that the conductance exhibits universal behavior. Namely we find that at large length scales, the conductance is independent of the hopping amplitude $t$ at the junction.

We also present in Sec.~\ref{section:Y_Junction} a numerical verification of a theoretical prediction for the conductance of the chiral fixed point. The chiral fixed point is stable in the range $1<g<3$ and in the presence of a time-reversal symmetry breaking flux $\phi \neq 0, \pi$.

\section {Conductance and correlations in a noninteracting lattice model} 
\label{section:Noninteracting_Conductance}
There is a well-established framework, namely the Landauer-B\"{u}ttiker formalism (see for example Ref.~\onlinecite{Nazarov09}), for calculating conductances of multi-wire quantum junctions in the absence of electron-electron interactions. This framework is applicable both in the continuum and to lattice models. The key quantities in this framework are the incoming and outgoing scattering states $\Psi_{\rm in}$ and $\Psi_{\rm out}$, which for a junction of $M$ wires can be represented by an $M\times1$ ($2M\times1$) column vector for spinless (spinful) electrons. The scattering states are labeled by momentum quantum number $k$, and the effect of the junction is encoded in an $M\times M$ ($2M\times 2M$) unitary scattering matrix for spinless (spinful) electrons. The scattering matrix relates the incoming and outgoing scattering states as 
\begin{equation}
\Psi_{\rm out}(k)=S(k)\Psi_{\rm in}(k).
\end{equation}

In the Landauer-B\"{u}ttiker formalism, conductance is simply related to the elements of the scattering matrix $S(k_F)$ at the Fermi level. Because including spin is a simple extension of the method developed in this paper, we work with spinless electrons here. The conductance between two different wires $G_{\alpha \beta}$ is then given by
\begin{equation}\label{eq:Landauer}
G_{\alpha\beta}=|S_{\alpha \beta}(k_F)|^2\: \frac{e^2}{h}.
\end{equation}

\begin{figure}[htb]
\centering
\includegraphics[width = 6 cm]{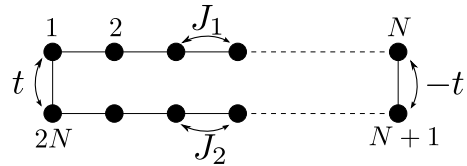}
\caption
 {A simple noninteracting system of two wires constructed with a junction with hopping $t$ and mirror image with hopping $-t$. The quantity of interest is the expectation value $\langle J^1(x)J^2(x) \rangle$ with $J^\alpha$ the current operator on wire $\alpha$.}
\label{fig:loop}
\end{figure}

Let us now consider the simplest junction of two lattice wires. Each wire has hopping amplitude set to unity in the bulk and the two wires are connected by a hopping amplitude $t$. More generically, one can consider a noninteracting junction of $M$ wires that are coupled quadratically as described below.
Consider a lattice system of $M$ wires with electron annihilation
operators $c_{\alpha,j}$, where $\alpha=1,...,M$ is the wire index and
$j=-\infty,...,0$ is the site index on each wire. We can write
the Hamiltonian in a compact form by defining
$\Psi_j^\dag=\left(c^\dag_{1,j}\;c^\dag_{2,j}\;  \cdots \;  c^\dag_{M,j} \right)$:
\begin{equation}\label{Hamiltonian}
{\cal H}=-\sum_{j=-\infty}^{-1}\left(\Psi_j^\dag
\Psi_{j+1}+\Psi_{j+1}^\dag \Psi_j\right) + \Psi_0^\dag \;\Gamma \;
\Psi_0,
\end{equation}
where $\Gamma$ is a Hermitian $M \times M$ matrix with diagonal
elements $\Gamma_{\alpha \alpha}=\mu^{\ }_{\alpha}$ (endpoint chemical potentials) and off-diagonal elements
$\Gamma_{\alpha\beta}=\Gamma^*_{\beta \alpha}=t_{\alpha \beta}$ (hopping between endpoints). In the bulk of each wire, the
nearest-neighbor hopping amplitude is set to unity. The scattering eigenstates are \begin{equation}|\psi\rangle_k=\sum_j
\Psi_j^\dag\left(A^+ e^{ikj}+A^- e^{-ikj} \right)|0\rangle,
 \end{equation}
where $\Psi^\dag_j$ is a row vector of operators $c^\dag_{\alpha, j}$ and  $A^{+(-)}$ is a column vector of scattering amplitudes $A^{+(-)}_\alpha$  for $\alpha=1\dots M$, and the standard matrix multiplication convention applies. By plugging the above states in the Schr\"{o}dinger equation ${\cal H}|\psi\rangle_k=\epsilon_k |\psi\rangle_k$, we obtain $\epsilon_k=-2\cos k$ and~\cite{Oshikawa06}
\begin{equation}\label{S}
    S(k)=-(\Gamma+e^{-ik})^{-1}(\Gamma+e^{ik})
\;.
\end{equation}
The scattering matrix relates the incoming and outgoing states $\Psi_{\rm in}=A^+$ and $\Psi_{\rm out}=A^-$ as $\Psi_{\rm out}=S\Psi_{\rm in}$. For the simple case of just two wires connected with hopping amplitude $t$, we have
\begin{equation}\label{eq:S_matrix}
 \Gamma=\left( \begin{matrix}
  0& -t \\
 -t & 0
        \end{matrix}\right),\qquad
S= \left( \begin{matrix}
  \frac{t^2-1}{t^2-e^{-2i k}}&\frac{t(e^{2i k}-1)}{e^{i k}t^2-e^{-i k}}  \\
 \frac{t(e^{2i k}-1)}{e^{i k}t^2-e^{-i k}}  & \frac{t^2-1}{t^2-e^{-2i k}}
        \end{matrix}\right).
\end{equation}
The above scattering matrix at half-filling then yields the following conductance :
\begin{equation}\label{eq:Landauer_conductance}
 G_{12}=\frac{4 t^2}{\left( 1+t^2\right) ^2}\frac{e^2}{h}.
\end{equation}
As discussed in the Introduction, the key result of this paper is a general relationship between the conductance and certain static correlation functions in a finite system. The purpose of this section is to illustrate this relationship in the very special case of a simple noninteracting system where exact calculations can be done with elementary methods. 
Consider the following finite system shown in Fig.~\ref{fig:loop}, which consists of our junction with hopping $t$ on the left. At a finite length away from the junction, the other endpoints of the wires are coupled with a hopping amplitude $-t$. Let us now calculate the current-current correlation function $\langle J^1(x)J^2(x) \rangle$.

\begin{figure}[htb]
\centering
\includegraphics[width= 8 cm]{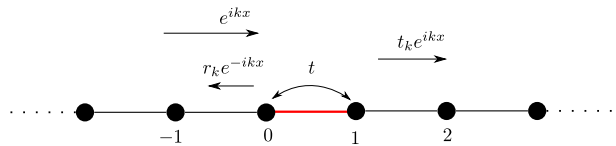}
\caption
 {(Color online) The single-particle scattering states for a single-impurity infinitely long system.}
\label{fig:noninteracting_semi_infinite}
\end{figure}

To begin with, let us assume a simple special case where $t=1$. In this case, our system is a loop, with no impurity, that is threaded with a flux $\pi$. This gives rise to anti-periodic BC. In this special case, the Hamiltonian can be simply diagonalized and we get
 \begin{equation}
 H=\sum_k -2 \cos k\: c^\dagger_k c_k, \quad k=\frac{2n-1}{2N}\pi
\end{equation}
where $c_k$ is the Fourier transform of $c_j$ with $j=1,\dots 2N$ and $n=-N+1, \dots N$. We focus on the half-filled case and assume $N$ is even. The ground state is then given by a Slater determinant $|{\rm GS}\rangle=c^\dagger_{k_1}\dots c^\dagger_{k_N}|0\rangle$ and the fermionic correlation function
\[
 C(a,b)\equiv \langle  {\rm GS}| c^\dagger_{a}c_{b} |{\rm GS} \rangle,
\]
where $a$ and $b$ are the site index as in Fig.~\ref{fig:noninteracting_semi_infinite}, can be written as
 \begin{equation}
C(a,b)=\frac{1}{2N}\sum_{n=1}^{N \over 2}\left[ e^{i \pi \frac{2n-1}{2N} (a-b)} +c.c. \right]. 
\end{equation}

Each term in the above sum corresponds to one filled momentum level where the contributions of $k$ and $-k$ levels are complex conjugates. The sum is just a geometric series, which can be calculated exactly and the result is 
 \begin{equation}\label{eq:fermionic_greens_function}
C(a,b)=\frac{1}{2N}\frac{\sin \frac{\pi (a-b)}{2}}{\sin \frac{\pi (a-b)}{2N}}. 
\end{equation}

If $J^1(m)$ (on wire $1$) is the current operator between sites $m$ and $m+1$, $J^2(m)$ will be between sites $2N-m$ and $2N-m+1$ in the chain. Let us write these operators explicitly as
\begin{eqnarray*}
J^1(m)&=&i\left(c^\dagger_m\: c_{m+1}-c^\dagger_{m+1}\: c_{m}\right),\\
J^2(m)&=&i\left(c^\dagger_{2 N+1-m}\: c_{2N-m}-c^\dagger_{2N-m}\: c_{2N-m+1}\right).
\end{eqnarray*}
Using the fermionic Green's function Eq.~(\ref{eq:fermionic_greens_function}) and Wick's theorem, we can explicitly calculate the current-current correlation function $\langle J^1(m) J^2(m) \rangle$. The real-space form of Wick's theorem can be  generically written in the form
\begin{equation}\label{eq:wicks_theorem}
\langle c^\dagger_m\: c^\dagger_n \:c_i \:c_j\rangle=
\langle c^\dagger_m\:c_j\rangle
\langle c^\dagger_n\:c_i\rangle
-
\langle c^\dagger_m\:c_i\rangle
\langle c^\dagger_n\:c_j\rangle.
\end{equation}

To calculate $\langle J^1(m) J^2 (m)\rangle$, we write it as a sum of four quartic terms and reduce each term to a sum of products of single-electron Green's functions. Notice that $C(a,b)=C(a-b)$, a function of the distance $a-b$ alone. After some algebra, we can then write the current-current correlation function as 
\begin{eqnarray}\label{eq:JJ_from_C}
\langle J^1(m) J^2 (m)\rangle&=&
2C^2(2N-2m)\\&-&2C(2N-2m+1)C(2N-2m-1)\nonumber.
\end{eqnarray}
By plugging the explicit form of $C(a-b)$ from Eq.~(\ref{eq:fermionic_greens_function}) into Eq.~(\ref{eq:JJ_from_C}), we then obtain the following exact expression for the current-current correlation function:
\begin{widetext}
\[
\langle J^1(m) J^2(m) \rangle=\frac{1}{2N^2}
\left\{
{\sin^2 \left[\pi (N-m)\right] \over \sin^2 \left[\frac{\pi}{N} (N-m)\right]}-
{\sin \left[\pi (N-m)+{\pi \over 2}\right] \sin \left[\pi (N-m)-{\pi \over 2}\right]
\over
 \sin \left[{\pi \over N} (N-m)+{\pi \over 2N}\right] \sin \left[{\pi \over n}(N-m)-{\pi \over 2N}\right]}
\right\}.
\]
\end{widetext}
Consider the above correlation function away from the two endpoints $N-m\gg 1$ and $m\gg 1$. We can then approximate the denominator of the second term as $\sin^2 \left[\frac{\pi}{N} (N-m)\right]$ and write
\begin{equation}\label{eq:no_impurity_corr}
\langle J^1(m) J^2(m) \rangle=
{1\over 2}\left[ N \sin \left(\pi {m \over N}\right) \right]^{-2}.
\end{equation}
As we will show later, the expression derived above for a simple noninteracting model has a universal form that survives electron-electron interactions. 

Before proceeding, let us consider a limit of the above expression. If we take the limit of $N\rightarrow\infty$ in Eq.~(\ref{eq:no_impurity_corr}) above, we are sending off the right junction to infinity and effectively describing a semi-infinite system. In this case, for a distance $m$ (lattice spacing set to unity), we obtain the following correlation function:
\begin{equation}\label{eq:no_impurity_corr_limit}
\langle J^1 (m)J^2(m) \rangle=
{1\over 2\pi^2}\frac{1}{m^2}.
\end{equation}
 
In the next step, we derive a similar expression for a semi-infinite system but with an impurity consisting of a site with hopping amplitude $t$. It is illuminating to first give a short derivation of Eq.~(\ref{eq:no_impurity_corr_limit}) formulated directly in the limit of the semi-infinite system. In this limit, we can treat the momenta in the continuum and work directly with the scattering wave functions, which in the special case of $t=1$ are just plane waves $e^{i k x}$ due to the absence of back-scattering. We have right-moving (left-moving) plane waves for $0<k$ ($0>k$) and at half-filling, the filled momentum states have $-\frac{\pi}{2}<k<\frac{\pi}{2}$. This leads to the following fermion Green's function:
\begin{equation}\label{eq:Greens_function1}
C(a,b)=\int_{-\frac{\pi}{2}}^{\frac{\pi}{2}}\frac{dk}{ 2 \pi}\: e^{ika}e^{-ikb}=\frac{\sin{\frac{\pi(a-b)}{2}}}{\pi (a-b)},
\end{equation}
which is in fact the limit of $N\rightarrow\infty$ of Eq.~(\ref{eq:fermionic_greens_function}). Inserting the above expression into
\[
 \langle J^1(m) J^2(m) \rangle=2C^2(a-b)-2C(a-b+1)C(a-b-1),
\]
where $m=(b-a)/2$ straightforwardly leads to Eq.~(\ref{eq:no_impurity_corr_limit}).

So far, we have only considered $t=1$. Let us now consider a semi-infinite system shown in Fig.~\ref{fig:noninteracting_semi_infinite} with an impurity of arbitrary hopping amplitude $t$.
Now, instead of the simple plane waves $e^{ikx}$, we can use the following single-particle scattering states. Let us write the right-moving and left-moving scattering states separately for clarity. We have

\begin{displaymath}
\phi_k^R(x)=\left\lbrace \begin{array}{l}
e^{ikx}+r_k e^{-ikx}, \quad x\leqslant 0 \\ 
t_k e^{ikx}, \qquad  ~~~~~~~~ x\geqslant 1\\ 
\end{array}\right.
\end{displaymath}

\begin{displaymath}
~~~\phi_k^L(x)=\left\lbrace \begin{array}{l}
e^{-ikx}+\tilde{r}_k e^{ikx}, \quad x\leqslant 0 \\ 
\tilde{t}_k e^{-ikx}, \qquad  ~~~~~~  x\geqslant 1.\\ 
\end{array}\right.
\end{displaymath}

In terms of the above scattering states, we can write the fermionic Green's function for $a<0$ and $b>1$ as
\begin{eqnarray}\label{eq:Greens_function2}
C(a,b)& = &\int_{0}^{\frac{\pi}{2}}\frac{dk}{ 2 \pi}\:\left[\left(e^{-ika}+r^*_k e^{ika} \right)t_k e^{ikb}\right. \nonumber\\  & & \left. +\tilde{t}^*_k e^{ika}\left(e^{-ikb}+\tilde{r}_k e^{ikb} \right)\right].  
\end{eqnarray}
Note that in the simple case $t=1$, where we have no back-scattering and $r_k=\tilde{r}_k=0$, Eq.~(\ref{eq:Greens_function2}) simply reduces to Eq.~(\ref{eq:Greens_function1}) above. By using the scattering matrix Eq.~(\ref{eq:S_matrix}) or more directly from plugging in the scattering states into the equations
\begin{eqnarray}
 \epsilon_k \phi_k(1)&=&-t\phi_k(0)-\phi_k(2),   \\
\epsilon_k \phi_k(0)&=&-t\phi_k(1)-\phi_k(-1),  
\end{eqnarray}
we obtain $r_k=\frac{t^2-1}{e^{-2ik}-t^2}$ and $t_k=t\frac{e^{-2ik}-1}{e^{-2ik}-t^2}$.
We also have $\tilde{t}_k= t_k$ and $\tilde{r}_k= e^{-2ik}r_k$.
By inserting the above transmission and reflection coefficients into the expression for the Green's function Eq.~(\ref{eq:Greens_function2}) and after some algebra, we can write
\begin{widetext}
\begin{equation}\label{eq:Greens_function3}
C(a,b)=4t\int_{0}^{\frac{\pi}{2}}\frac{dk}{ 2 \pi}\:\sin k \:\frac{ \sin\left[ k(a-b+1)\right] t^2-\sin\left[ k(a-b-1)\right] }{1-2t^2 \cos 2k +t^4}\equiv C(a-b).  
\end{equation}
\end{widetext}

\begin{figure}[htb]
\centering
\includegraphics[width= 8 cm]{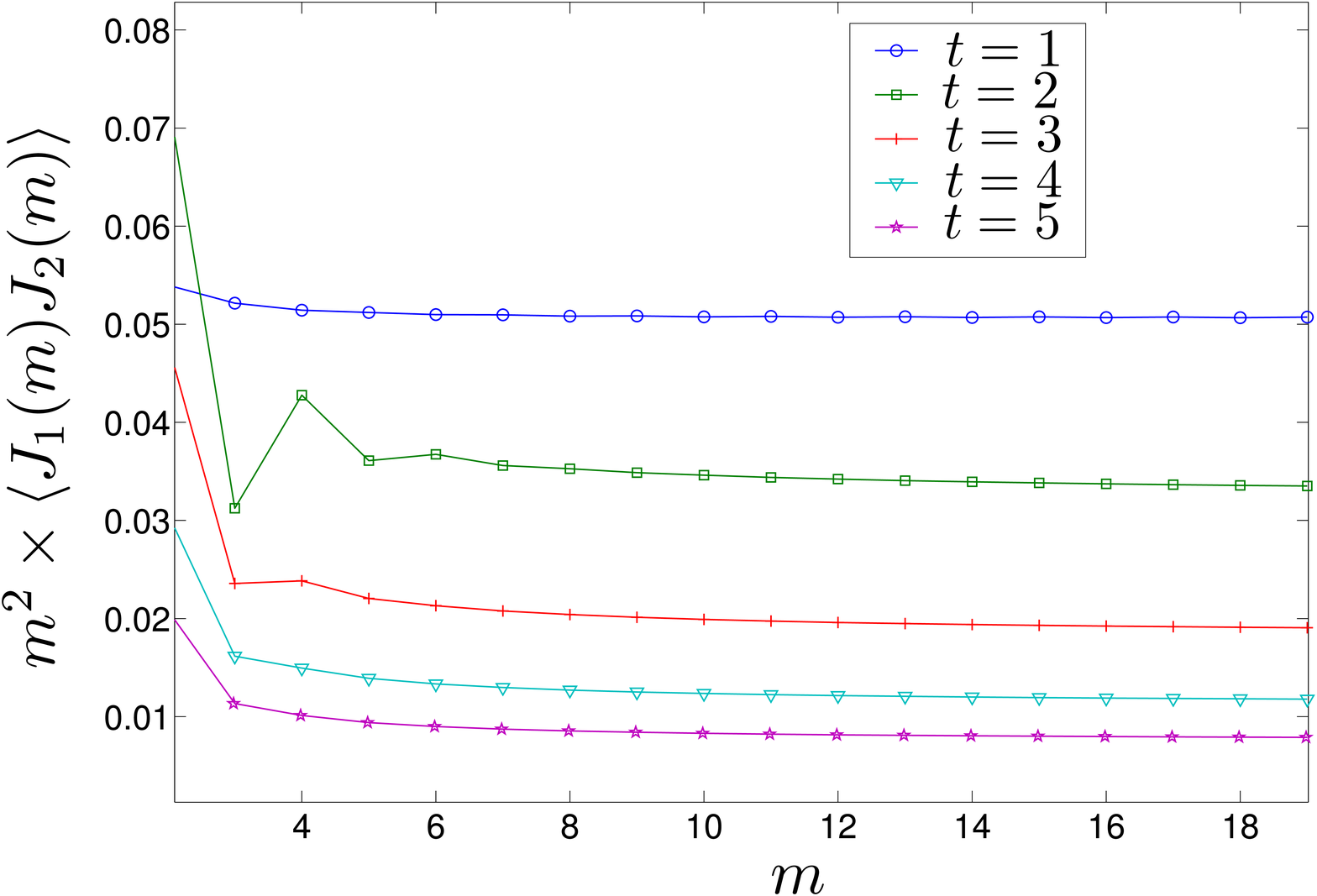}
\caption
 {(Color online) The quantity $m^2\langle J^1 (m)J^2(m) \rangle $ calculated from Eq.~(\ref{eq:J1J2}) by a numerical integration of the integral in Eq.~(\ref{eq:Greens_function3}). For large $m$, this quantity saturates to a constant value that only depends on the hopping amplitude $t$. This indicates that the asymptotic form of $\langle J^1(m) J^2(m) \rangle $ is given by $\sim \frac{1}{m^2}$.}
\label{fig:JJ_corr}
\end{figure}

This leads to the following expression for the current-current correlation function:
 \begin{equation}\label{eq:J1J2}
\langle J^1(m) J^2(m) \rangle =2C^2(-2m)-2C(-2m-1)C(-2m+1)
\end{equation}
with $C(x)$ defined in Eq.~(\ref{eq:Greens_function3}). We claim that the asymptotic behavior of the quantity $\langle J^1(m) J^2(m) \rangle$ is generically given by $\sim \frac{1}{m^2}$ as in the special case Eq.~(\ref{eq:no_impurity_corr_limit}). To check this claim we plot $m^2\langle J^1(m) J^2(m) \rangle$ as a function of $m$ for several values of the hopping amplitude $t$ by straightforward numerical evaluation of the integral in Eq.~(\ref{eq:Greens_function3}). The results are shown in Fig.~\ref{fig:JJ_corr}.

Since for each $t$, $m^2\langle J^1(m) J^2(m) \rangle $ saturates to a constant, a natural question is how this constant depends on the value of $t$. A quite remarkable fact is that the saturation value is exactly proportional to the conductance Eq.~(\ref{eq:Landauer_conductance}) of this simple junction, which we calculated in the beginning of this section from the Landauer's formalism. In other words, we have the following asymptotic behavior:
\begin{equation}\label{eq:infinite:scaling}
 \langle J^1 (m)J^2 (m)\rangle \simeq \frac{1}{2 \pi^2}\frac{4 t^2}{\left( 1+t^2\right) ^2}\frac{1}{m^2}, \qquad m\gg1.
\end{equation}

At this level, this observation appears rather mysterious. It is not very transparent from the expression for $\langle J^1(m) J^2(m) \rangle$, which is a rather complicated double integral, why this ground state correlation function has such a simple scaling form. By comparing Eq.~(\ref{eq:infinite:scaling}) above with Eq.~(\ref{eq:no_impurity_corr}), we claim that for a finite system shown in Fig.~\ref{fig:loop}, the asymptotic behavior of the correlation function is given by
\begin{equation}\label{eq:noninteracting_key}
 \langle J^1(m) J^2(m) \rangle\simeq \frac{2 t^2}{\left( 1+t^2\right) ^2}\left[ N \sin \left(\pi {m \over N}\right) \right]^{-2}, \qquad m\gg1.
\end{equation}

We do not attempt to prove the above expression here. The equation is indeed a very special case of the generic relationship Eq.~(\ref{eq:key_relation_intro}) that we prove in section~\ref{section:Conductance_Correlations} in full generality, i.e., in the presence of electron-electron interactions for a rather arbitrary junction with an arbitrary number of wires and without assuming symmetries such as time-reversal.

 The expression above, however, motivates the main idea of this paper, i.e., the fact that conductance can be extracted from ground state expectation values in a closed system, in a very elementary example. It is also an example where the relationship can be derived explicitly on the lattice instead of resorting to the continuum formalism, which helps clarify the application of the continuum CFT results to a lattice calculation.

\section{Review of bosonization and boundary conformal field theory}
\label{section:Bosonization_CFT}
The main results of this paper are derived within the framework of boundary conformal field theory. In this section, we review the required steps for formulating the generic system we would like to study, namely a junction of $M$ quantum wires modeled as a tight-binding Hamiltonian with spinless electrons, in the language of the conformal field theory. The material in this section can be skipped by readers familiar with bosonization and CFT. 
\subsection{Bosonization of quantum wires}
\label{subsection:Bosonization}
\begin{figure}[htb]
\centering
\includegraphics[width = 6 cm]{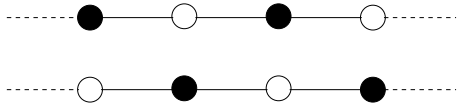}
\caption
 {The degenerate ground states of  the Hamiltonian Eq.~(\ref{eq:one-wire}) for  $V\rightarrow\infty$. Filled circles represent occupied sites.}
\label{fig:CDW}
\end{figure}

The first step in this formulation is the bosonization procedure. Bosonization is a powerful nonlocal transformation that, in one space dimension, allows us to describe the low-energy limit of strongly-correlated fermionic systems as a noninteracting theory of bosons (see, for example, Ref.~\onlinecite{Senechal99} for a review). Let us start by considering one infinitely long quantum wire with the following Hamiltonian:
 \begin{equation}\label{eq:one-wire}
 H=\sum_i\left[
  -c^\dagger_i c_{i+1} - c^\dagger_{i+1} c_{i}+V(n_i-\frac{1}{2})(n_{i+1}-\frac{1}{2})\right], 
\end{equation}
where $n_i=  c^\dagger_i c_{i}$. The first two terms describe electron hopping and the last term is the density-density interaction. At half-filling, this model exhibits a charge density wave phase transition for large repulsive interactions $V>2$. We then see that the ground state spontaneously breaks lattice translation symmetry and we get the two degenerate ground states shown in Fig.~\ref{fig:CDW} for $V\rightarrow\infty$.

Also at very large attractive interactions, the electrons will clump together and form clusters of neighboring occupied sites. This results in phase separation which, at half-filling, happens at $V<-2$.

For a certain range of interactions $-2<V<2$, which includes the noninteracting fermionic system ($V=0$), the system is in a gapless critical phase known as the Luttinger liquid. This is the regime where bosonization applies. Let us first bosonize the noninteracting system.

The noninteracting fermions have an $\epsilon_k=-2 \cos (k)$ dispersion and a ground state consisting of a filled Fermi sea up to the Fermi level $k_F$. We can linearize the dispersion around the Fermi level and only consider the right-moving (left-moving) excitations close to $k_F$ ($-k_F$). 

Let us define the following right-moving and left-moving fields:
 \begin{equation}\label{eq:chiral_fermion}
\psi_R(p)=c(k_F+p), \qquad \psi_L(p)=c(-k_F+p).
\end{equation}
We can then write the hopping part of the Hamiltonian as 
\begin{equation}
 H_0=v_F \int_{-\Lambda}^{\Lambda}\frac{d p}{2\pi}
\left[p \psi^\dagger_R(p)\psi_R(p)-p\psi^\dagger_L(p)\psi_L(p) \right]. 
\end{equation}
Upon Fourier transforming, the above equation in real space reads as
\begin{equation}\label{eq:H_0}
 H_0=v_F \int d x
\left[\psi^\dagger_R(x)\:\frac{1}{i}\frac{\partial}{\partial x}\:\psi_R(x)
-\psi^\dagger_L(x)\:\frac{1}{i}\frac{\partial}{\partial x}\:\psi_L(x) \right],
\end{equation}
where
\begin{equation}
\psi(x)=e^{i k_F x}\psi_R(x)+e^{-i k_F x}\psi_L(x).
\end{equation}

The key step to bosonization is writing the Hamiltonian (both hopping and interaction part) in terms of the following chiral current operators:
\begin{equation}\label{eq:chiral_current}
j_{R,L}(x)=\psi^\dagger_{R,L}(x)\:\psi_{R,L}(x)
\end{equation}
instead of the fermionic creation and annihilation operators. One can easily verify using the Hamiltonian $H_0$ in Eq.~(\ref{eq:H_0}) that these currents satisfy the following commutation relations:
\begin{equation}\label{eq:chiral_current_H_0_commutations}
\left[H_0,j_{R,L}\right]=\pm i v_F \frac{\partial}{\partial x}j_{R,L}.\end{equation}

The interaction term, which is quartic in creation and annihilation operators, will trivially become quadratic in terms of the currents. The nontrivial part of the bosonization procedure is to show that the hopping part of the Hamiltonian, which is quadratic in $\psi$, will remain quadratic in terms of $j$.

This important result can be shown using the commutation relation of the current operators. By Fourier transforming $\psi_R$ as $\psi_R(x)=\frac{1}{\sqrt{N}}\sum_ke^{i k x}\psi_R(k)$, we can write the Fourier transform of the chiral current $j_R$ in Eq.~(\ref{eq:chiral_current}) as
\begin{equation}\label{eq:chiral_current_fourier}
j_{R}(q)=\frac{1}{\sqrt{N}}\sum_k\psi^\dagger_R(k-q)\psi_R(k),
\end{equation}
which leads to the following commutation relation~\cite{Haldane81,Mahan}:
\begin{equation}\label{eq:chiral_current_commutations}
\left[j_R(q),j_R(q^\prime)\right]=\frac{q}{2\pi}\delta_{q,-q^\prime}. 
\end{equation}
A similar equation with $\frac{q}{2\pi}\rightarrow-\frac{q}{2\pi}$ can be derived for $j_L$. Using the commutation relation Eq.~(\ref{eq:chiral_current_commutations}), one can show that the bosonized Hamiltonian 
\begin{equation}\label{eq:H_0-bosonized}
H_0=\pi v_F \sum_q\left[j_R(q) j_R(-q) + j_L(q) j_L(-q)\right]
\end{equation} 
obeys the commutation relations in Eq.~(\ref{eq:chiral_current_H_0_commutations}). We then argue that  Hamiltonian (\ref{eq:H_0-bosonized}) is the bosonized form of the noninteracting fermionic Hamiltonian (\ref{eq:H_0}).

By using the commutation relations between $j_{R,L}$ and $\psi_{R,L}$ , one can write the fermionic operators in terms of the currents by introducing bosonic fields $\phi_{R,L}$. The result is
\begin{equation}\label{eq:bosonization_dictionary}
\psi_{R,L}=\frac{1}{\sqrt{2\pi}}e^{\mp i \phi_{R,L}},\qquad j_{R,L}=\frac{1}{2\pi}\frac{\partial}{\partial x}\phi_{R,L}.
\end{equation}
The interacting Hamiltonian density will now have diagonal terms of the form $j^2_L(x)$, $j^2_R(x)$ coming from the hopping and off-diagonal terms of the form  $j_L(x)j_R(x)$ from the electron-electron interaction (back-scattering) and is quadratic in terms of the currents. By perfroming a linear transformation 
\begin{equation}\label{eq:chiral_rotation}
\left(\begin{matrix}
 J_R\\ 
J_L
\end{matrix}
 \right)=
\left(
\begin{array}{cc}
 \cosh \beta & \sinh  \beta \\ 
 \sinh\beta & \cosh \beta
\end{array} 
\right)
\left(\begin{matrix}
 j_R\\ 
j_L
\end{matrix}
 \right)
\end{equation}
that preserves the commutation relations, we obtain
\begin{equation}
{\cal H}(x)\propto\left[ J^2_R(x)+J^2_L(x)\right].
\end{equation}

It is convenient to define new bosonic fields $\varphi$ and $\theta$ that are linear combinations of $\phi_{R,L}$ in Eq.~(\ref{eq:bosonization_dictionary}) such that 
\begin{equation}
 \psi_{R,L}(x) = \frac{1}{\sqrt{2\pi} } e^{i (\varphi(x) \pm \theta(x))/\sqrt{2} }.
\end{equation}
In terms of these fields and by introducing the renormalized charge carrier velocity $v$ and the dimensionless Luttinger parameter $g$, the low energy effective Hamiltonian density for a wire can be written as 
\begin{equation}
\label{eq:boson-H}
{\cal H}(x)= \frac{v}{4\pi} \left[ g (\partial_x \varphi )^2 +\frac{1}{g} (\partial_x \theta)^2 \right].
\end{equation}

The bosonized Hamiltonian Eq.~(\ref{eq:boson-H}) is the generic description of 1D metallic quantum wires that we work with in this paper. Notice that $\frac{1}{2\pi}\partial_x \varphi$ is the momentum conjugate to $\theta$ and we have the following commutation relations:
\begin{equation}
[ \varphi(x), \theta(x')] = i \pi {\rm sgn}(x' - x). 
\end{equation}
The above review covers the main ingredients of the bosonization scheme that we need for setting up the problem at hand in this paper.

\subsection{Boundary conformal field theory}
\label{subsection:BCFT}
Here, we briefly review the basics of CFT and BCFT that we need in the remainder of this paper. This review is largely based on Ref.~\onlinecite{Cardy10}.
An important property of critical theories, i.e., theories  where certain fields known as the scaling fields have critical correlation functions, is scale invariance. This simply means that the correlation functions of scaling fields $O_i$ satisfy
\begin{eqnarray}
&&\langle O_1 (b\: r_1)O_2 ( b\: r_2)\dots O_n ( b\: r_n)\rangle\nonumber \\
&& ~~~~~=b^{-\sum_i x_i} \langle O_1 (r_1)O_2 (  r_2)\dots O_n ( r_n)\rangle,
\end{eqnarray}
where the exponent $x_i$ is known as the scaling dimension of the operator $O_i$ and $b$ is an arbitrary scaling factor.

A powerful leap from scaling symmetry to conformal symmetry is by allowing the scaling factor $b$ to vary smoothly, i.e., considering more general transformation $r\rightarrow r^\prime$ than $r\rightarrow br$ with $b(r)=|\frac{\partial r^\prime}{\partial r}|$. If we have a local theory and the transformation $r\rightarrow r^\prime$ locally resembles a scaling transformation (modulo a local rotation) then we expect \begin{eqnarray}\label{eq:conformal_symm}
&&\langle O_1 (r^\prime_1)O_2 (  r^\prime_2)\dots O_n (  r^\prime_n)\rangle \nonumber \\
&&~~~~~=\prod_i b(r_i)^{- x_i} \langle O_1 (r_1)O_2 (  r_2)\dots O_n ( r_n)\rangle.
\end{eqnarray}
In three dimensions, the transformations that locally resemble scaling are limited but in two dimensions (2D) (or 1+1D quantum systems) any analytic mapping [$z\rightarrow w(z)$ on the complex plane $z$] is conformal (preserves angles) and requiring conformal symmetry leads to highly nontrivial results. An important quantity in a CFT is the stress-energy tensor. Consider the action $S$ of a 1+1D quantum system that is defined on the complex plane $z$. The stress-energy tensor $T_{\mu \nu}$ is defined in terms of the variations of this action through
\begin{equation}
\delta S=-\dfrac{1}{2 \pi}\int T_{\mu \nu} \partial_\nu\alpha^{\mu} d^2 r,
\end{equation}
where $\delta S$ is the variation of $S$ due to the infinitesimal transformation $r^\mu \rightarrow r^\mu+\alpha^\mu(r)$. 
Another important ingredient of CFT is the operator product expansion (OPE), which describes the nature of the singularities in $\langle O_i(r_i) O_j(r_j)\dots\rangle$ as $r_i\rightarrow r_j$ as 
\begin{equation}
\langle O_i(r_i)O_j(r_j) \dots\rangle=\sum_k C_{ijk}(r_i-r_j)\left\langle O_k(\frac{r_i+r_j}{2})\dots\right\rangle.
\end{equation}
In any CFT, the \textit{primary} fields are fields for which the most singular term in the OPE of $T(z) O_j(z_j,\bar{z}_j)$ is order $(z-z_j)^{-2}$ where $T(z)\equiv T_{zz}$.

Let us consider primary fields with the following scaling transformation on the complex plane $z$:
\[
O_j(\lambda z,\bar{\lambda}\bar{z})=\lambda^{-\Delta_j}\bar{\lambda}^{-\bar{\Delta}_j}O_j(z,\bar{z}),
\]
where $\Delta_j$ and $\bar{\Delta}_j$ are called the complex scaling dimensions.
Notice that the above equation is only a shorthand to describe the scaling behavior of correlation functions involving $O$. An important result in CFT is that under a generic conformal transformation $z\rightarrow w(z)$, the correlation functions of primary fields transform as~\cite{Cardy10}
\begin{eqnarray}\label{eq:primary_transformation}
&&\langle O_1(z_1,\bar{z}_1) O_2( z_2 , \bar{z}_2) \dots \rangle \nonumber\\
&&~~~~~=\prod_i \left|\frac{dw_i}{d z_i}\right|^{\Delta_i}\left |\frac{d \bar{w}_i}{d \bar{z}_i}\right|^{\bar{\Delta}_i} \langle O_1(w_1,\bar{w}_1) \dots \rangle.
\end{eqnarray}
This is in fact the same as Eq.~(\ref{eq:conformal_symm}) which we intuitively wrote down in the beginning of this section. We shall emphasize that the transformation \eqref{eq:primary_transformation} only holds for the primary fields of the theory.



So far our discussion has been limited to CFTs on the entire complex plane $z$. A CFT on a domain with boundaries is known as boundary conformal field theory (BCFT) where in addition to the bulk properties, we need to specify appropriate boundary conditions (BCs). An important task in BCFT is classifying the conformally invariant boundary conditions for a given bulk CFT.~\cite{Cardy89}

The BCFT techniques have proved powerful in studying various quantum impurity problems such as the Kondo model and Luttinger liquids with impurities.~\cite{Wong94,Affleck10} The effect of the impurity, in this approach to quantum impurity problems, is to select the appropriate conformally invariant BC corresponding to the low-energy RG fixed point. At far away from the boundaries, the bulk CFT is expected to adequately describe the system, i.e., the boundary conditions do not matter. Also very close to the boundary, there is nonuniversal short-distance physics from the microscopic degrees of freedom. On distances from the boundary that are much larger than the microscopic length scales, but still much smaller than the domain size, the correlation functions are affected by the conformally invariant boundary conditions.~\cite{Cardy10} 

\begin{figure}[htb]
\centering
\includegraphics[width= 6 cm]{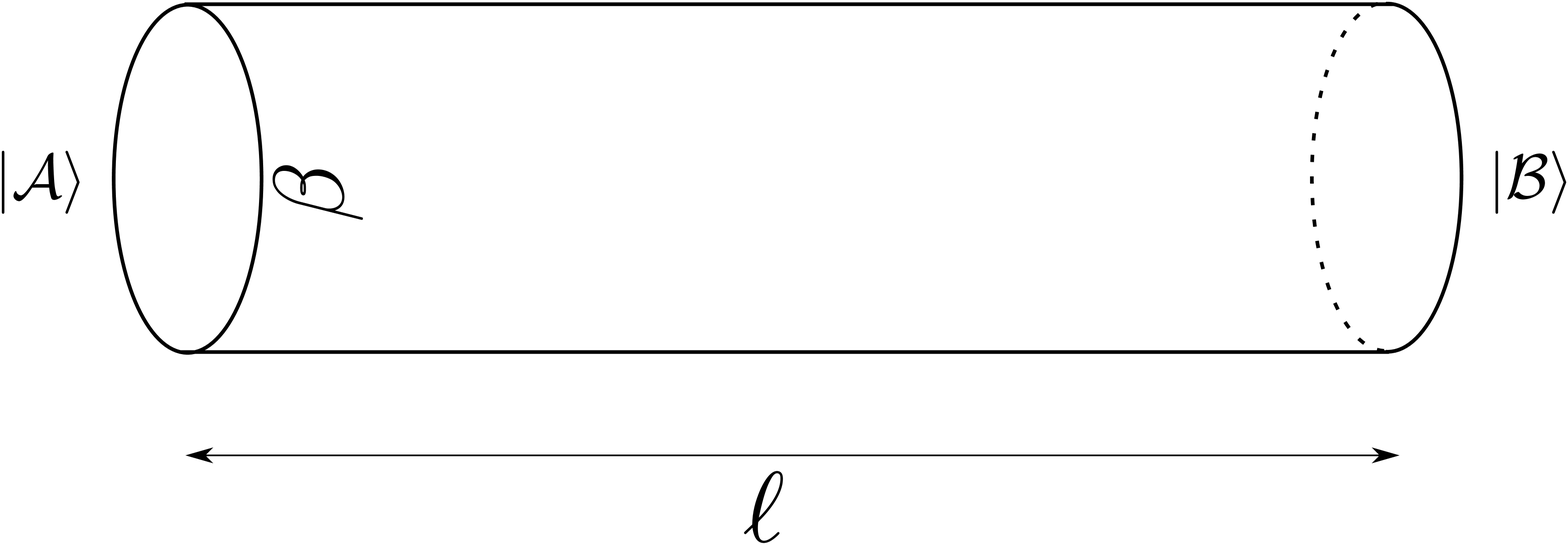}
\caption
 {Boundary states in the partition function of a CFT on a cylinder. }
\label{fig:cylinder}
\end{figure}

As pointed out by Cardy, the BCs in a BCFT are encoded in boundary states.~\cite{Cardy89} To understand the notion of a boundary state, it is helpful to consider the partition function of a CFT on a cylinder with boundary conditions $\cal{A}$ and $\cal{B}$ on two sides of the cylinder as in Fig.~\ref{fig:cylinder}.~\cite{Saleur98} One can write down the partition function using transfer matrices running parallel or perpendicular to the boundary.  
 The transfer matrix in the direction parallel to the boundary depends on the boundary conditions ${\cal A}$ and $\cal{B}$ and can be written as $\exp\left( {-H_{\cal{AB}}}\right)$ (the lattice spacing is set to unity). The imaginary time runs parallel to the boundary and $H_{\cal{AB}}$ is the Hamiltonian with boundary conditions ${\cal A}$ and $\cal{B}$. We then have the partition function
\begin{equation}\label{eq:Z}
 {\cal Z} = {\rm tr}\left[ e^ {-\beta \:H_{\cal{AB}}}\right]. 
\end{equation}
Alternatively one can construct the transfer matrix in the direction perpendicular to the boundary. The Hamiltonian $H$ for this transfer matrix is determined by the bulk only and is independent of the boundary conditions. To write the partition function we need to construct states $|{\cal A}\rangle$ and $|\cal{B}\rangle$ in the space where the transfer matrix acts such that the partition function Eq.~(\ref{eq:Z}) is also equal to
\begin{equation}\label{eq:Z-BS}
 {\cal Z} = \langle {\cal A}| e^ {-\ell\: H}|\cal{B}\rangle. 
\end{equation}
The two different representations of the partition function above illustrate the notion of the boundary state.~\cite{Cardy10}

\section{A general relationship between the conductance and static
correlation functions}
\label{section:Conductance_Correlations}

Let us recall the two difficulties with calculating the conductance of a junction of multiple interacting quantum wires, which we discussed in the Introduction. First, conductance is a property of an open quantum system. This means that we think of the wires emanating from the junction as being attached to reservoirs that can serve as sources and drains for the electrons passing through the molecular structure at the junction. To study the conductance, we either need to faithfully model the reservoirs in our theoretical description or assume that we have infinitely long wires. In the latter case, we basically send the reservoirs to infinity. An appropriate boundary condition is assumed at infinity, which does not enter the theory of the junction explicitly. For a numerical calculation of conductance, it then seems that we would need to model large enough systems that approximate the semi-infinite ones.

The second difficulty is that, in the standard linear-response framework, the linear conductance is related to dynamical rather than static correlation functions. This can be seen explicitly from the Kubo formula for the conductance tensor~\cite{Oshikawa06}
\begin{eqnarray}\label{eq:Kubo}
 &&G_{\alpha \beta} = \lim_{{\omega} \rightarrow 0_+}-\frac{e^2}{\hbar}\frac{1}{
	{\omega} L}\int_{-\infty}^\infty d\tau\;e^{i {\omega}\tau}\\
 && ~~~~~~~~~~~\times
	\int_0^L dx \;\langle {\cal T}_\tau J^\alpha(y,\tau) J^\beta(x,0)\rangle,\nonumber
\end{eqnarray}
where ${\cal T}_\tau$ indicates imaginary-time ordering and the quantity $\langle {\cal T}_\tau J^\alpha(y,\tau) J^\beta(x,0)\rangle$ is a dynamical current-current correlation function for currents $J^\alpha$ and $J^\beta$ on wires $\alpha$ and $\beta$, respectively. It seems that, generically, a numerical calculation of this quantity requires time-dependent methods such, as for example, time-dependent DMRG. 

With a brute force approach, the numerical calculation of conductance is not feasible for junctions of three or more wires. In this section, we derive a key relationship between the universal conductance and certain static correlation functions in a finite system. This important relationship is the basis of a method for numerically calculating the conductance that we will discuss in Sec.~\ref{section:Method}. 
\subsection{CFT for independent wires}
Let us begin by formulating the problem in the BCFT framework reviewed in Sec.~\ref{section:Bosonization_CFT}. We have $M$ quantum wires connected to a junction. Let us first consider the system of $M$ independent quantum wires in the absence of the junction. As argued in Sec.~\ref{subsection:Bosonization}, the effective Luttinger-liquid Hamiltonian for one wire can be written as
\begin{equation}
H= \frac{v}{4\pi} \int d x \left[ g (\partial_x \varphi )^2 +\frac{1}{g} (\partial_x \theta)^2 \right].
\end{equation}  

Using the fact that $\varPi_\theta=\frac{1}{2 \pi}\partial_x \varphi$ is the conjugate momentum of $\theta$, the Lagrangian density can be written as
\[
{\cal L}=\frac{1}{4 \pi g}\left[\frac{1}{v} (\dot{\theta})^2-v\left(\partial_x \theta\right)^2\right]. 
\]
Let us set the charge carrier velocity $v$ to unity for convenience. We can then write the Euclidean action as 
\begin{equation}
\label{eq:Luttinger_action}
S=\frac{1}{ 4 \pi g} \int d\tau d x  \, \partial_\mu \theta \partial^\mu \theta. 
\end{equation} 
Alternatively the action can be written in terms of the dual field $\varphi$ as $S={g \over 4 \pi} \int d\tau d x \, \partial_\mu \varphi \partial^\mu \varphi$. It is convenient to represent the points on the $(x,\tau)$ plane with complex coordinate $z=\tau+i x$. The system described by the massless action Eq.~(\ref{eq:Luttinger_action}) is a CFT on the complex plane $z$. Physically, covering the whole complex plane implies 1D quantum wires extending from $-\infty$ to $\infty$ at zero temperature ($\beta\rightarrow\infty$).

Let us begin by calculating the correlation functions of the bosonic fields $\theta$ from the action Eq.~(\ref{eq:Luttinger_action}).
We have 

\begin{eqnarray}
\langle {\cal T}_\tau \theta(x_1, \tau_1) \theta(x_2, \tau_2)\rangle&=&\frac{\int {\cal D} [\theta]\: \theta(x_1, \tau_1) \theta(x_2, \tau_2) e^{-S}}{\int {\cal D} [\theta]\: e^{-S}}\nonumber\\
&=&K(x_1-x_2,\tau_1-\tau_2),
\end{eqnarray}
where the propagator $K(x_1-x_2,\tau_1-\tau_2)$ is the inverse of the operator $-{1 \over 2 \pi g}\left(\partial_x^2 +\partial_\tau^2 \right) $, i.e.,
\begin{equation}
-{1 \over 2 \pi g}\left(\partial_x^2 +\partial_\tau^2 \right)K(x-x^\prime,\tau-\tau^\prime)=\delta(x-x^\prime)\delta(\tau-\tau^\prime)
\end{equation}
and is given by
\begin{equation}
K(x-x^\prime,\tau-\tau^\prime)=-\frac{g}{2}\ln\left[ (x-x^\prime)^2+(\tau-\tau^\prime)^2\right]+{\rm const.} 
\end{equation}
Up to an unimportant additive constant, we can then write the correlation function of the bosonic fields $\theta$ as 
\begin{equation}\label{eq:theta_correlation}
\langle {\cal T}_\tau \theta(z_1,\bar{z}_1) \theta(z_2,\bar{z}_2)\rangle=-\frac{g}{2}\left[\:\ln \left(z_1-z_2\right)+\ln\left(\bar{z}_1-\bar{z}_2\right)\:\right].
\end{equation}By differentiating the above Eq.~(\ref{eq:theta_correlation}) with respect to $z$ and $\bar{z}$, we obtain the following correlation functions:
\begin{eqnarray}
  \left\langle {\cal T}_\tau\: \frac{\partial \theta}{\partial z_1}(z_1,\bar{z}_1)\: \frac{\partial \theta}{\partial z_2}(z_2,\bar{z}_2)\right \rangle&=&
-\frac{g}{2}\frac{1}{(z_1-z_2)^2},
 \\
 \left\langle {\cal T}_\tau \:\frac{\partial \theta }{\partial \bar{z}_1}(z_1,\bar{z}_1)\: \frac{\partial \theta}{\partial \bar{z}_2}(z_2,\bar{z}_2)\right\rangle&=&
-\frac{g}{2}\frac{1}{(\bar{z}_1-\bar{z}_2)^2}.
\end{eqnarray}
Also we can similarly show that $\langle {\cal T}_\tau \:\frac{\partial \theta }{\partial z_1}(z_1,\bar{z}_1)\: \frac{\partial \theta}{\partial \bar{z}_2}(z_2,\bar{z}_2)\rangle$ vanishes.

The chiral currents below, where we have put in a wire index $\alpha$, are then primary fields for this CFT:
\begin{equation}
J^\alpha_L(z)=\frac{i}{\sqrt{2}\pi}\: \partial \: \theta^\alpha(z,\bar{z}), \quad J^\alpha_R(\bar{z})=-\frac{i}{\sqrt{2}\pi}\: \bar{\partial} \: \theta^\alpha(z,\bar{z}),
\end{equation}
with the notation
\[
\partial \equiv \partial_z={1 \over 2}(\partial_\tau-i\partial_x), \quad 
 \bar{\partial} \equiv
\partial_{\bar {z}}={1 \over 2}(\partial_\tau+i\partial_x).
\]

In the absence of a boundary, the chiral currents in different wires are uncorrelated and the only correlations are between chiral currents of the same chirality in the same wire. The only nonvanishing correlation functions of the chiral currents are then
\begin{equation}\label{eq:chiral_corr}
\begin{split}
  \langle {\cal T}_\tau J_L^\alpha(z_1)J_L^\beta(z_2)\rangle&= {g\over 4 \pi ^2}\frac{\delta_{\alpha \beta}}{(z_1-z_2)^2},\\
  \langle {\cal T}_\tau J_R^\alpha(\bar{z}_1)J_R^\beta(\bar{z}_2)\rangle&= {g\over 4 \pi ^2}\frac{\delta_{\alpha \beta}}{(\bar{z}_1-\bar{z}_2)^2}.
\end{split}
\end{equation}

Note that the fact that there is no correlation between the left-movers and the right-movers does not mean that there is no back-scattering in the system. Indeed, we actually have back-scattering in the presence of electron-electron interactions. The chiral currents $J_{L,R}$ are the chiral eigenmodes of the interacting theory which, as explained in Eq.~(\ref{eq:chiral_rotation}), are in fact linear combinations of the bare (noninteracting) chiral currents $j_{L,R}$. Interactions induce back-scattering for these bare currents $j_{L,R}$.

The form of the correlation functions above is highly constrained by conformal symmetry and is basically determined by the scaling dimension of the primary operators. A very important observation is that space and time are tied together in the above correlation functions and a measurement of static correlation functions uniquely determined the dynamical ones. Let us note that these chiral currents are related to the physical current $J$ and the density fluctuation $\rho$ through
\[
 J^\alpha=v(J_R^\alpha-J_L^\alpha), \quad  \rho^\alpha =J_R^\alpha +J_L^\alpha.
\]
The velocity of the charge carriers $v$ coincides with the Fermi velocity only in the absence of electron-electron interactions. For simplicity, we shall set $v$ to unity in the remainder of this section.
\subsection{BCFT for the semi-infinite system}

Let us now return to the problem of the quantum junction. The system of $M$ independent wires is described by a CFT of $M$ bosonic fields living on the full complex $z$ plane. The junction of the $M$ quantum wires at the RG fixed point is described by BCFT on the upper-half complex plane. The real axis $x=0$ is the boundary of the domain where the CFT lives. We expect to have the same bulk CFT as the system of $M$ independent wires if we are infinitely far away from this boundary.

This is an example of a quantum impurity problem. An important hypothesis, which has been repeatedly verified in a multitude of quantum impurity problems,~\cite{Affleck10} such as the single-channel and multichannel Kondo model,~\cite{Affleck91} is that at the RG fixed points, the conformal symmetry in the bulk terminates smoothly. More precisely, the hypothesis states that the BC on the boundary of the domain in the complex plane is conformally invariant and therefore the quantum impurity problem at the RG fixed point is described by a BCFT.

How can the presence of the boundary change the correlation functions? In other words, what is the difference between the correlation functions in the CFT on the full $z$ plane and the BCFT on the upper half plane? A key observation is that the presence of the boundary does not change the correlation functions between chiral currents of the same chirality. On physical grounds, it is easy to see that the left movers coming from infinity toward the junction have not yet felt the presence of the junction and, just by causality, can not have different correlation functions than in the system of $M$ independent wires. For the right movers, on the other hand, we need the conformal invariance of the boundary conditions to make such a statement.

If we have a conformally invariant BC on the real axis, $\langle J_R^\alpha(\bar{z}_1)J_R^\alpha(\bar{z}_2)\rangle$ will be the same as on the full complex plane. This is generically the case in the limit when the two points $z_1$ and $z_2$ are far away from the boundary and deep in the bulk of the system. With conformally invariant BCs, however, these points can be brought close to the boundary through conformal transformations, such as $z\rightarrow \lambda z$ or $z\rightarrow -1/z$, from the upper half-plane onto itself, which leaves $\langle J_R^\alpha(\bar{z}_1)J_R^\alpha(\bar{z}_2)\rangle$ invariant. The above argument is not a rigorous proof, but may give some intuition. Notice that irrelevant boundary operators can give corrections to $\langle J_R^\alpha(\bar{z}_1)J_R^\alpha(\bar{z}_2)\rangle$ in the semi-infinite system, which decay as a power law with the distance from the boundary with exponents larger than $2$. In a finite system, both $\langle J_L^\alpha(z_1)J_L^\alpha(z_2)\rangle$ and $\langle J_R^\alpha(\bar{z}_1)J_R^\alpha(\bar{z}_2)\rangle$ will have such corrections.

The presence of a boundary, however, does introduce new correlations between the left and the right movers. From the scattering picture, we can understand this in the sense that left movers must go through the junction before turning into right movers. The form of the new correlation functions is determined by the scaling dimension of the current operators, and as far as these correlation functions are concerned, all the information about the boundary condition at $x=0$ is encoded in an overall coefficient $A_{\cal B}^{\alpha \beta}$. 

In summary, in the BCFT describing the junction at the RG fixed point, in addition to the correlation functions Eq.~(\ref{eq:chiral_corr}), we have the following nonvanishing correlation functions~\cite{Wong94}:
\begin{equation}\label{eq:right_left_corr}
 \langle {\cal T}_\tau J_R^\alpha(\bar{z}_1)J_L^\beta(z_2)\rangle= -{g\over 4 \pi ^2}A_{\cal B}^{\alpha \beta}\frac{1}{(\bar{z}_1-z_2)^2} \; 
\end{equation}
and a similar equation for  $\langle {\cal T}_\tau J_L^\alpha(z_1)J_R^\beta(\bar{z}_2)\rangle$.

 The coefficients $A_{\cal B}^{\alpha \beta}$ are determined by the conformally invariant boundary condition on the real axis. In fact we can write this coefficient in terms of the boundary state $|\cal B\rangle$ as~\cite{Wong94,Cardy_Lewellen91}
\begin{equation}\label{eq:highest_weight}
 {g\over 4 \pi ^2}A_{\cal B}^{\alpha \beta}=\frac{\langle J_R^\alpha J_L^\beta,0|{\cal
B}\rangle}{\langle 0|{\cal B}\rangle},
\end{equation}
where $|O,0\rangle$ is the highest weight state corresponding to a generic operator $O$ (here $O=J^\alpha_L J^\beta_R$) and $|0\rangle$ is the ground state. The definition of the highest weight state and a derivation of Eq.~(\ref{eq:highest_weight}) is given in appendix~\ref{app:highest_weight}. 

One important observation is in order. While the Hamiltonian is local in terms of the fermionic degrees of freedom, the mapping from fermion to bosons, i.e, the bosonization procedure, is highly nonlocal. One must notice, however, that regardless of the number of wires, there is a fundamental difference between a two-dimensional system and this quasi-1D system. The nonlocality only shows up along the boundary itself. The statistics of the microscopic degrees of freedom can affect what the boundary condition ${\cal B}$, and consequently the coefficient $A_{\cal B}^{\alpha \beta}$, is but it does not change the fact that the effect of the junction can be reduced to a BC at $x=0$. The generic form Eq.~(\ref{eq:right_left_corr}) holds even though the bosonic fields are nonlocal objects in terms of the original fermions.

 So far, we have established that the current-current correlation functions appearing in Eq.~(\ref{eq:Kubo}) have a universal form modulo coefficients $A_{\cal B}^{\alpha \beta}$, which depend on the boundary conditions. We should then be able to express the conductance $G_{\alpha \beta}$ in terms of these coefficients $A_{\cal B}^{\alpha \beta}$ by performing the integrals in the Kubo formula. Consider the conductance $G_{\alpha \beta}$ between two distinct wires $\alpha\neq \beta$. Using $J=J_R-J_L$ (velocity set to unity), we can write
\begin{eqnarray*}
&&\langle {\cal T}_\tau J^\alpha(z_1, \bar{z}_1) J^\beta(z_2, \bar{z}_2)\rangle\\
&&~~~~~~={g\over 4 \pi ^2}\left[ A_{\cal B}^{\alpha \beta}\:\frac{1}{(\bar{z}_1-z_2)^2}+A_{\cal B}^{\beta \alpha}\:\frac{1}{(z_1-\bar{z}_2)^2}\right].
\end{eqnarray*}

Note that the presence of the boundary does not change the correlation between currents of the same chirality, and the chiral correlation functions are obviously equal to zero in the half-plane for $\alpha \neq \beta$ (because they are proportional to $\delta_{\alpha \beta}$). Plugging the above equation into the Kubo formula Eq.~(\ref{eq:Kubo}) (with $z_1=\tau +iy$ and  $z_2=ix$) gives two similar terms. Let us first perform the integral over the imaginary time $\tau$. We can write
\begin{equation}
\int_{-\infty}^{+\infty} d \tau\:\frac{e^{i \omega \tau}}{\left( \tau-iu\right) ^2}=-2\pi \omega H(u)e^{-\omega u}, \qquad u\neq0
\end{equation}
which can be easily derived by a contour integration. Here $H(u)$ is the Heaviside step function $H(u)=1(0)$ for $u>0(u<0)$. Using the above identity we can then write
\begin{eqnarray*}\label{eq:Kubo2}
G_{\alpha \beta}&=& \lim_{\epsilon \rightarrow 0}\: g \frac{e^2}{ h}\frac{1}{L}\\
	 & &\int_\epsilon^L dx \left[ A^{\alpha \beta}_{\cal B} \;H(x+y)+A^{\beta \alpha}_{\cal B}\; H(-x-y)\right],
\end{eqnarray*}
which simply yields
\begin{equation}\label{eq:GA}
G_{\alpha \beta}=A^{\alpha \beta}_{\cal B} \; g\frac{e^2}{ h}, \quad \alpha  \neq \beta.
\end{equation}
In the above expression, since both $x$ and $y$ are positive (the wire extends from $x=0$ to $+\infty$), the second term identically vanishes. Notice that determining the off-diagonal elements $G_{\alpha \beta}$ with $\alpha \neq \beta$ is enough to also determine the diagonal elements of the conductance tensor $G$. The elements are not independent and satisfy the relations
\begin{equation}
\label{eq:G-conservation-relation}
\sum_ \beta G_{\alpha \beta}=0, \quad \sum_\alpha G_{\alpha \beta}=0.
\end{equation}
The first relation follows because applying the same voltage to all the wires leads to zero current, and the second relation follows from current conservation. 

Notice that in deriving the above expression~(\ref{eq:GA}), we assumed that the interacting Luttinger-liquid wires extend to infinity. In many experimental situations, the interacting wires are attached to noninteracting Fermi liquid leads. If the interacting region is long enough, the effect of the Fermi-liquid leads can be taken into account by considering a contact resistance as discussed later in Sec.~\ref{section:Y_Junction}. Short interacting wires attached to Fermi-liquid leads can be considered as part of the junction, while leads constitute quantum wires with $g=1$, which extend to infinity. The possible crossover behavior as a function of the length of the interacting region in the presence of infinite noninteracting leads is an interesting open question.

One of the two difficulties with a numerical calculation of the universal conductance is now effectively solved. The conductance is uniquely determined by the coefficient in front of the dynamical correlation function Eq.~(\ref{eq:right_left_corr}). But, conformal symmetry determines the universal form of this correlation function. In other words, space and time are tied together by conformal symmetry, and the coefficient $A^{\alpha \beta}_{\cal B}$ can be extracted from a static correlation function alone. More explicitly, consider the static correlation function $\langle J_R^\alpha(x)J_L^\beta (x)\rangle$ that is merely a special case of the generic correlation function $\langle {\cal T}_\tau J_R^\alpha (\bar{z_1})J_L^\beta (z_2)\rangle$ in Eq.~(\ref{eq:right_left_corr}) for $z_1=z_2=i x$ and is given by~\cite{Wong94}
\begin{eqnarray}\label{eq:right_left_static}
\langle  J_R^\alpha (x)J_L^\beta(x)\rangle= {g\over 4 \pi ^2}A_{\cal B}^{\alpha \beta}\frac{1}{(2x)^2}. 
 \end{eqnarray}
The above correlation function is just a ground-state expectation value in a semi-infinite system. If we can numerically compute this ground-state expectation value and, as a function of the distance $x$ from the boundary, fit it to the power-law form above, we can extract the coefficient $A_{\cal B}^{\alpha \beta}$ and uniquely determine the conductance.

Because the BCFT description applies only at low energies and large distances, the behavior of the system at short distances (in units of the lattice spacing) is nonuniversal and depends on the microscopic details rather than the continuum limit. Even at distances where the continuum limit is valid, the expression above only gives the asymptotic behavior of the correlation function. The conformally invariant boundary condition $\cal B$ describes the RG fixed point, but, generically, there are corrections to the BCFT prediction Eq.~(\ref{eq:right_left_static}) from irrelevant boundary operators at finite distances. These corrections of course die  out faster than $\frac{1}{x^2}$, and at large distances the asymptotic behavior is given by the leading term Eq.~(\ref{eq:right_left_static}).

\subsection{Conformal mapping from the semi-infinite to a finite system}

Having solved one of the difficulties of calculating the conductance, we now turn to the other difficulty. Namely, the fact that modeling an open, i.e., semi-infinite, system requires finding correlation functions in very large systems. Let us go back to the BCFT description of the semi-infinite junction for which the conductance is well defined. We argued that this system is described by a BCFT on the upper half-plane. As explained below, we can use a conformal mapping to go from the upper half-plane to a strip of width $\ell$, which describes a finite system of length $\ell$ at zero temperature. We will show in what follows that the correlation function $\langle  J_R^\alpha(x)J_L^\beta(x)\rangle$ in this finite system also has a universal form and allows the extraction of the key coefficient $A_{\cal B}^{\alpha \beta}$.

\begin{figure}[htb]
\centering
\includegraphics[width = 8 cm]{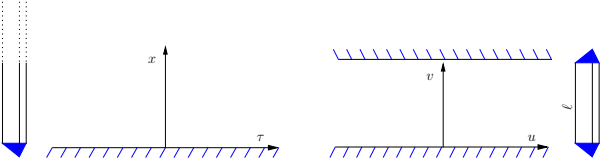}
\caption
{(Color online) The conformal transformation from the upper half-plane to a strip of width $\ell$. The physical system described by the half-plane is a semi-infinite junction at zero temperature with the imaginary time running from $-\infty$ to $+\infty$ and the position along the wires running from $0$ to $+\infty$. The physical system corresponding to the strip is a finite system of length $\ell$ at zero temperature. The structure of the junctions at the two end-points of this finite system must give rise to the appropriate boundary conditions consistent with the conformal mapping.}
\label{fig:mapping}
\end{figure}

Let us first explain the mapping, shown in Fig.~\ref{fig:mapping}, from the upper half-plane $z=\tau+i x$ to a strip $w=u+iv$. The mapping is given by
\begin{equation}\label{eq:conformal_transformation}
 w= \frac{\ell}{\pi} \ln z, \qquad z=e^{\frac{\pi}{\ell} w}.
\end{equation}
In polar coordinates we can write $z=r e^{i \phi}$ on the half-plane with $0\leq r\leq\infty$ and  $0 \leq \phi \leq \pi$ which gives $u=\frac{\ell}{\pi} \ln r$ and $v=\frac{\ell}{\pi} \phi$. The $x>0$ part of the real axis in the $z$ plane is mapped to the boundary $v=0$ and the $x<0$ part to $v=\ell$. 

In the finite system of length $\ell$, which we obtain from the conformal mapping above, the correlation functions behave in a universal manner. 
Let us formally consider the static correlation function of Eq.~(\ref{eq:right_left_static}), which is only a function of the distance from the boundary $x$, as a function of $z=\tau+ix$ but with no dependence on $\tau$. If we have a primary operator $O$ (here $O=J_R^\alpha J_L^\beta$) in the semi-infinite plane  with
\begin{equation}\langle O(z) \rangle=A_{\cal B}^O(2x)^{-X_O},\end{equation}
we can use the expression Eq.~(\ref{eq:primary_transformation}) to write the expectation value of the same operator in the strip as 
\begin{equation}
\langle O(w) \rangle=|{dw \over dz}|^{-X_O}\langle O(z) \rangle=A_{\cal B}^O\left(2\frac{\ell}{\pi}\frac{x}{|z|} \right)^{-X_O}. 
\end{equation}
Using $|z|=e^{\frac{\pi}{2\ell}(w+w^*)}=e^{\frac{\pi}{\ell}u}$ and $x=\textrm{Im}\: e^{\frac{\pi}{\ell}w}=e^{\frac{\pi}{\ell}u} \sin\left( \frac{\pi}{\ell}v\right) $, we obtain
\begin{equation}
\langle O(w) \rangle=\left[2\: \frac{\ell}{\pi} \sin\left( \frac{\pi}{\ell}v\right)  \right]^{-X_O}, 
\end{equation}
where $v=\textrm{Im}(w)$ is physically the distance from the left boundary ($v=0$) in the finite system (see the right-hand side of Fig.~\ref{fig:mapping}). Since the zero-temperature finite system is invariant under
translations in $u$, the static ground-state expectation value of the
local operator $J_R^\alpha(x)J_L^\beta(x)$, in the finite system of length $\ell$ obtained from the transformation Eq.~(\ref{eq:conformal_transformation}), is expected to behave as
\begin{equation}\label{eq:key_eq1}
 \langle J_R^\alpha(x)J_L^\beta(x) \rangle_{\rm GS}={g\over 4 \pi ^2}A_{\cal B}^{\alpha \beta} \left[2\:
{\sin\left( {\pi \over \ell}x\right) }/{{\pi \over \ell}}
\right]^{-2}
\end{equation}
with $x$ the distance from aboundary.

A simple check of the above expression is that it reproduces, as expected, the correct power-law expectation value of the semi-infinite system Eq.~(\ref{eq:right_left_static}) in the limit $\ell\rightarrow\infty$. By combining the above expression with the result of Eq.~(\ref{eq:Kubo2}), we obtain 
\begin{equation}\label{eq:key_relation}
 G_{\alpha \beta}=\langle J_R^\alpha(x)J_L^\beta(x) \rangle_{\rm GS}\left[4\: \ell\:
\sin\left( {\pi \over \ell}x\right) 
\right]^{2} \frac{e^2}{h}.
\end{equation}

Notice that as long as the correlation function in the above expression is calculated for large enough $x$, i.e., larger than the microscopic length scales and the required healing length for the contributions of the irrelevant boundary operators to become negligible, the value of $G_{\alpha \beta}$ should be independent of $x$. Practically, one can extract $G_{\alpha \beta}$ by fitting the calculated $\langle J_R^\alpha(x)J_L^\beta(x) \rangle_{\rm GS}$ to the form of the universal sine function. Notice that the assumption that all wires have the same Luttinger parameter $g$ was not essential for the derivation of Eq.~(\ref{eq:key_relation}). This is because the generic form of Eq.~(\ref{eq:right_left_corr}) holds even if the wires $\alpha$ and $\beta$ have different Luttinger parameters. The overall coefficient of $\frac{1}{(\bar{z}_1-z_2)^2}$ however will depend on the two Luttinger parameters as well as the boundary condition $\cal B$.  

The Eq.~(\ref{eq:key_relation}) is one of the key results of this paper. It is significant both from a fundamental and a practical point of view. From a fundamental viewpoint, we think about conductance when we have a state with currents flowing through the system from a source and into a drain. It is remarkable that one can construct a closed finite system, the ground state of which fully encodes the conductance, a quantity defined in open systems. From the practical point of view, the above expression is the basis for a method of calculating the universal conductance that we discuss in the next section where we show how to apply this continuum result to a lattice computation and how to implement the boundary conditions.

\section{A method for calculating the conductance}
\label{section:Method}

In the previous section, we derived a general expression that related the off-diagonal elements $G_{\alpha \beta}$ of the conductance tensor to the static correlation functions $\langle J^\alpha_R(x) J^\beta_L(x) \rangle$. The key to this relation is that both the conductance and the static correlation function above are proportional to only one coefficient that depends on the conformally invariant boundary condition at the junction. This relationship can be used as the basis of a method for calculating the conductance.

If we measure, numerically, the static correlation function $\langle J^\alpha_R(x) J^\beta_L(x) \rangle$ in a finite system obtained from the conformal transformation Eq.~(\ref{eq:conformal_transformation}), we can then fit the correlation function (a function of $x$) to the universal form Eq.~(\ref{eq:key_eq1}) and obtain the coefficient $A_{\cal B}^{\alpha \beta}$. It is important to note that the universal expression is the asymptotic form of the correlation function and the fit should be done in a region far away from the boundary. We need to consider large enough $x$ so that the microscopic details become unimportant and the corrections due to irrelevant operators become negligible.

To carry out this scheme, we need to compute $\langle J^\alpha_R(x) J^\beta_L(x) \rangle$ in the ground state. The computation is amenable to the time-independent DMRG method. In the quasi-1D system obtained by having all wires running parallel to one another, this correlation function is actually the expectation value of a \textit{local} operator, which makes it a particularly easy quantity to calculate with DMRG. We need to answer two questions, however, before we can proceed. 
\begin{enumerate}
\item Since we cannot model chiral creation and annihilation operators on a lattice, how can we model the operator $J^\alpha_R(x) J^\beta_L(x)$?
\item What exactly is the finite system that we need to put into a computer? In other words, given a tight-binding Hamiltonian for the original semi-infinite system, what is the Hamiltonian of the finite system for which the correlation function $\langle J^\alpha_R(x) J^\beta_L(x) \rangle$ has the universal form we derived in the continuum? 
\end{enumerate}

To answer the first question, we make use of the following \textit{continuum} relationship between the chiral currents and the total current and the total charge density:
\begin{equation}\label{eq:J_and_ N}
 J^\alpha=v(J_R^\alpha-J_L^\alpha), \quad  N^\alpha=J_R^\alpha+J_L^\alpha.
\end{equation}
Notice that since with electron-electron interactions these chiral currents are not the bare left and right-moving currents, the velocity $v$ is not the Fermi velocity but the renormalized velocity of the charge carriers that can be calculated from the Bethe ansatz~\cite{Giamarchi03} and is given by the following expression at half-filling:
\begin{equation}\label{eq:Bethe_velocity}
v=\pi\frac{\sqrt{1-(V/2)^2}}{\arccos\:(V/2)},
\end{equation}
where we have set the hopping amplitude to unity and $V$ is the interaction strength in the bulk lattice Hamiltonian Eq.~(\ref{eq:one-wire}). From the Bethe ansatz, we can also obtain the effective Luttinger parameter in terms of the microscopic lattice Hamiltonian and, at half-filling, we have
\begin{equation}\label{eq:Bethe_g}
g=\frac{\pi}{2\arccos\:(-V/2)}.
\end{equation}

Can anything go wrong if we apply the continuum relation~(\ref{eq:J_and_ N}) to a lattice computation? The relation Eq.~(\ref{eq:J_and_ N}) is only exact in the continuum. However, we find that it gives very accurate results on the lattice for correlation functions between different wires $\alpha \neq \beta$. We will see this explicitly in the numerical studies of the following sections. The simple lattice calculations we performed in Sec.~\ref{section:Noninteracting_Conductance} also shed some light on this issue.

These off-diagonal correlation functions are those that we actually need for the conductance calculation, as in our approach, we first find the off-diagonal elements of the conductance tensor, which are simply proportional to $A^{\alpha \beta}_{\cal B}$. The diagonal ones can be later deduced from Eq.~\eqref{eq:G-conservation-relation}. Now, we would like to find the off-diagonal elements of the conductance tensor $G_{\alpha \beta}$, and we use Eq.~(\ref{eq:J_and_ N}) as if it were an operator identity. Using Eq.~(\ref{eq:J_and_ N}), we can then
write, for $\alpha \neq \beta$,
\begin{eqnarray*}
  \langle J^\alpha(x)J^\beta(x)\rangle=&-v^2\left(  \langle J_L^\alpha(x)J_R^\beta(x)\rangle+\langle J_R^\alpha(x)J_L^\beta(x)\rangle\right),\nonumber \\ 
 \langle N^\alpha(x)J^\beta(x)\rangle=&v\left(  \langle J_L^\alpha(x)J_R^\beta(x)\rangle-\langle J_R^\alpha(x)J_L^\beta(x)\rangle\right) .
\end{eqnarray*}
The correlation function $\langle J_R^\alpha(x)J_L^\beta(x)\rangle$ is then simply given by
\begin{equation}\label{eq:measure}
\langle J_R^\alpha (x)J_L^\beta(x)\rangle=-\dfrac{1}{2v_\alpha v_\beta}\langle J^\alpha(x)J^\beta (x)\rangle-\dfrac{1}{2v_\beta}\langle N^\alpha(x)J^\beta(x)\rangle,
\end{equation}
where $v_\alpha$ ($v_\beta$) is the velocity in wire $\alpha$ ($\beta$). Throughout this paper, we work with wires with the same Luttinger parameter and therefore $v_\alpha=v_\beta=v$ but the more general expression above can be used in cases where the wires have different Luttinger parameters.

In writing the above expressions, we have made use of the fact that the chiral correlation functions of the form $\langle J_R^\alpha (x)J_R^\beta(x)\rangle$ vanish in this finite system for $\alpha\neq \beta$. We argued in section~\ref{section:Conductance_Correlations} that these correlation functions vanish in the semi-infinite system (upper half-plane). Since the finite system corresponds to the strip obtained from the upper half-plane by a conformal transformation, they also vanish in the finite system. 

Note that if we have time-reversal symmetry, as in the case of the simple two-wire problem studied in section \ref{section:Noninteracting_Conductance}, the second term in the above equation vanishes. This is because (nonchiral) currents $J$ are odd under time-reversal and the densities $N$ are even. If the structure at the junction does not include magnetic fluxes, time-reversal symmetry remains unbroken and we only need to calculate one current-current correlation function.

The operators $J$ and $N$ are easily modeled in terms of the lattice creation and annihilation operators. Two important observations are in order. First, since the current operator is defined for a bond, we have to write an effective bond density so the current and density are associated with the same position $x$. This is done by taking the average of the two site densities. The second observation is that bosonization is done for the charge density fluctuations and the background charge should be subtracted to construct the operator $N$. With these two observations, we can explicitly write
\begin{eqnarray*}
  J^\alpha (m+\dfrac{1}{2})&=&i({c}^\dagger_{\alpha,m+1}c_{\alpha,m}-{c}^\dagger_{\alpha,m}c_{\alpha,m+1}), \\ 
N^\alpha(m+\dfrac{1}{2})&=&\frac{1}{2}\left (n_{\alpha,m}+n_{\alpha,m+1}- \langle n_{\alpha,m}\rangle - \langle n_{\alpha,m+1}\rangle\right) .
\end{eqnarray*}

\begin{figure}[htb]
\centering
\includegraphics[width= 8 cm]{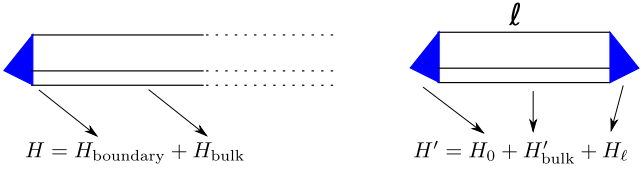}
\caption
{(Color online) Given the bulk and boundary Hamiltonians of a semi-infinite system, the goal is to obtain the two boundary Hamiltonians $H_0$ and $H_\ell$ of a finite system corresponding to the conformal transformation to the strip. }
\label{fig:finite_vs-infinite}
\end{figure}

We now address the question of implementing the boundary conditions. To measure the expectation values above, we need the ground state of a finite system. What we need to do is write a lattice Hamiltonian that corresponds to the continuum system we obtained from a conformal mapping. As discussed in the Introduction, the starting point is the Hamiltonian of the junction, which generically can be written as 
\[
H=H_{\rm boundary} +H_{\rm bulk}.
\]
The finite system has two boundaries on the left and on the right and as shown in Fig.~\ref{fig:finite_vs-infinite}, generically, a Hamiltonian
\begin{equation}
H^\prime=H_0 +H^\prime_{\rm bulk} +H_\ell.
\end{equation}

Since the bulk theory is conformally invariant, we can write $H^\prime_{\rm bulk}$ by using exactly the same terms as in $H_{\rm bulk}$, but just a finite number of them. The more subtle issue is to identify the boundary Hamiltonians $H_0$ and $H_\ell$ that give the correct conformally invariant boundary conditions. There are some details regarding persistent currents and the parity of the number of electrons that we discuss later, but for now let us give a generic argument regarding the boundary conditions.

The semi-infinite system that corresponds to the upper-half plane has one boundary and a conformally invariant boundary condition on this boundary, i.e., the real axis. Now, a conformally invariant boundary condition, by definition, does not change under a conformal transformation. The strip on the $w=u+iv$ plane, however, has two boundaries at $v=0$ and $v=\ell$. Let us review the conformal transformation $w=\frac{\ell}{\pi} \ln z$. Writing $z$ in polar coordinates as $z=r e^{i \phi}$ gives $v=\frac{\ell}{\pi} \phi$, so the left boundary comes from the positive real axis ($\phi=0$) and the right boundary from the negative real axis ($\phi=\pi$). Therefore, both these boundaries are expected to have the same boundary condition as the real axis in the semi-infinite system. 

Now, in the semi-infinite system, this unknown conformally invariant boundary condition is stabilized by the structure and interactions of the junction. That is to say a \textit{cap} with Hamiltonian $H_{\rm boundary}$ gives rise to the boundary condition $\cal B$ at $x=0$. It is then a completely natural assumption that the same cap should give the boundary condition $\cal B$ at the left boundary of the finite system. This simply yields
\[
H_0=H_{\rm boundary}.
\]

The boundary Hamiltonian on the other side ($H_\ell$), however, is generically different than $H_0$. Let us for convenience represent the distance from the left boundary by $x$ in the finite system as well as in the semi-infinite one. As argued above, we wish to have the same boundary condition on both sides. This does \textit{not} mean, however, that we should place the same cap on both ends of the finite system. To understand the issue, let us consider the case of noninteracting electrons, where the boundary condition can be simply written in the form of a single-particle scattering matrix.

\begin{figure}[htb]
\centering
\includegraphics[width= 8 cm]{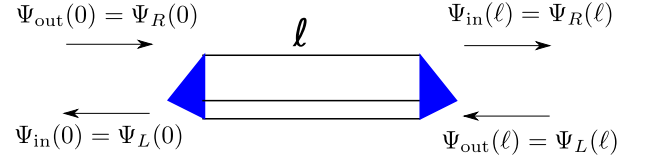}
\caption
{(Color online) A generic finite system with two boundaries. For noninteracting fermions, the boundary conditions can be written in the form of scattering matrices that relate the in-coming and out-going states. The correspondence of these states with the chiral fermions is illustrated at both boundaries.}
\label{fig:left_right_cap}
\end{figure}

Suppose the $S$ matrix of the left boundary ($x=0$) is $S_0$. As seen in Fig.~\ref{fig:left_right_cap}, the incoming scattering states at $x=0$ are the right-moving fermions $\psi_R(0)$ and the out-going states are the left-moving fermions $\psi_L(0)$. The boundary condition can then be written in terms of $S_0$ as 
\begin{equation}\label{eq:left_BC}
\psi_R(0)=S_0\:\psi_L(0).
\end{equation}   
As evident from Fig.~\ref{fig:left_right_cap}, the left and the right movers switch roles at the two boundaries. Namely, the left movers  (right movers)  are the out-going (in-coming) scattering states for the right boundary and the in-coming (out-going) scattering states for the left boundary. If the scattering matrix corresponding to the Hamiltonian of the right cap $H_\ell$ is $S_\ell$, we have by definition
\begin{equation}\label{eq:right_BC1}
\psi_L(\ell)=S_\ell\:\psi_R(\ell).
\end{equation} 

However, we want to have the same boundary \textit{condition} at the two boundaries. This means that to write the boundary condition of the right boundary at $x=\ell$, we should write exactly the same expression as Eq.~(\ref{eq:left_BC}) but replace the argument $x=0$ with $x=\ell$, which yields   
\begin{equation}\label{eq:right_BC2}
\psi_L(\ell)=S_0^{-1}\:\psi_R(\ell)=S_0^{\dagger}\:\psi_R(\ell).
\end{equation}
Comparing Eqs.~(\ref{eq:right_BC1}) and (\ref{eq:right_BC2}) leads to the simple but important equation
\begin{equation}\label{eq:S_R_S_L}
S_\ell=S_0^{-1}=S_0^{\dagger}.
\end{equation}

Still working with a noninteracting system with the boundary condition~(\ref{eq:left_BC}) for the semi-infinite system, let us derive the boundary conditions of the finite system more carefully. The chiral fermionic creation and annihilation operators are in fact primary fields with scaling dimension $\dfrac{1}{2}$ and transform as 
\begin{equation}\label{eq:fermion_xform}
\Psi_L(w)=\left(\frac{dw}{dz} \right)^{-\frac{1}{2}}\Psi_L(z), \quad \Psi_R(\bar{w})=\left(\frac{d\bar{w}}{d\bar{z}} \right)^{-\frac{1}{2}}\Psi_R(\bar{z}).
\end{equation}
By using the derivatives 
\[
\frac{dw}{dz}={ \ell \over \pi} \dfrac{1}{z}={ \ell \over \pi} e^{-{\pi \over \ell}w},
 \qquad
  \frac{d\bar{w}}{d\bar{z}}={ \ell \over \pi} \dfrac{1}{\bar{z}}={ \ell \over \pi} e^{-{\pi \over \ell}\bar{w}},
\]
we find that, at the two endpoints of the finite system,
\begin{eqnarray}
\frac{d\bar{w}}{d\bar{z}}&=&\frac{dw}{dz}{\big |}_{v=0}=\frac{\ell}{\pi}e^{-\frac{\pi}{\ell}u},\\
\frac{d\bar{w}}{d\bar{z}}&=&\frac{dw}{dz}{\big |}_{v=\ell}=-\frac{\ell}{\pi}e^{-\frac{\pi}{\ell}u}.
\end{eqnarray}

The minus sign above indicates that there could be subtle issues with the assumption of having the same boundary condition at both ends, which lead us to write Eq.~(\ref{eq:right_BC2}). Of course, notice that the minus-sign difference between the transformed chiral fermions at the two endpoints of the finite system is not inconsistent with Eq.~(\ref{eq:right_BC2}) because we also need to take the square root of $\frac{dw}{dz}$ and $\frac{d\bar{w}}{d\bar{z}}$ as seen from Eq.~(\ref{eq:fermion_xform}). At this level, it is not clear, generically, which branch of the square root we should pick, but the two choices lead to the following two possibilities for the scattering matrix at the right endpoint:
\[
{\rm I:}\quad S_\ell=S^{-1}_0,\qquad{\rm II:}\quad S_\ell=-S^{-1}_0. 
\]
 
The above relations can be thought of as two \textit{microscopic} BCs, which in conjunction with other microscopic details such as the number of electrons in the system give rise to the correct BC in the continuum. Although we can microscopically implement both of these scattering matrices, we work with the relation I above in this paper. Our approach is to use the relation I and choose the number of electrons such that, as discussed below, no persistent currents are generated in the finite system with this microscopic BC. 

The presence of persistent currents can give corrections to the CFT predictions. The reason persistent currents need to be avoided is that the best region for fitting the correlation function $\langle J_R^\alpha(x) J_L^\beta(x) \rangle$ is away from the boundaries and close to the center of the finite system where the effect of the irrelevant boundary operators is negligibly small. In the center of the finite system, the correlation function will be ${\cal O}(\dfrac{1}{\ell^2})$ and since persistent currents go as $\dfrac{1}{\ell}$, the corrections to $\langle J_R^\alpha(x) J_L^\beta(x) \rangle$, which are not accounted for in our formulation, become of the same $\dfrac{1}{\ell^2}$ order as the value of the $\langle J_R^\alpha(x) J_L^\beta(x) \rangle$ correlation function.

As an example for how to avoid persistent currents consider a two-wire system with the boundary condition I, i.e., the system shown in Fig.~\ref{fig:loop} of Sec.~\ref{section:Noninteracting_Conductance}. We find that the persistent currents are avoided by having an even number of electrons. This can be seen by noting that for hopping amplitude at the junction $t=1$ (a simple loop with a $\pi$ flux and no actual impurity), the allowed momentum levels exclude $k=0$  and are symmetric around it. Therefore, filling in an odd number of electrons leads to degeneracies.

Having gained some intuition regarding the boundary conditions in the absence of electron-electron interactions, we now tackle the problem of constructing the Hamiltonian $H_\ell$ of the right cap from that of the left, i.e., $H_0=H_{\rm boundary}$. We do this in two steps described below.

First, consider the case where the interactions are only in the wires and $H_{\rm boundary}$ is quadratic. In this case we imagine turning off the bulk interactions. We then \textit{engineer} a Hamiltonian $H_\ell$ in a completely noninteracting system, which gives the scattering matrix $S_\ell=S_0^{\dagger}$ at the desired filling. We now have a microscopic system with the \textit{same} boundary conditions at both endpoints. Turning on the interactions causes the system to flow away from the original noninteracting fixed point. The boundary conditions at both endpoints also flow away. However, since they were the same initially, we expect them to flow \textit{together} and reach a new BC. The two boundaries are then expected to have the same BC even with interactions.

The next step is to find symmetry transformations that automatically generate the desired Hamiltonian $H_\ell$ from the given $H_0$. In what follows we argue that at half-filling, a combination of time-reversal and particle-hole transformations does the job. Having the boundary conditions implemented with such simple transformations strongly suggests that these transformation should also implement the correct BCs even when we have interactions in the junction itself as well as in the bulk of the wires.

Our starting point for this procedure is an explicit expression for the scattering matrix from Sec. \ref{section:Noninteracting_Conductance}, which we repeat here for convenience: 
\begin{equation}
    S(k)=-(\Gamma+e^{-ik})^{-1}(\Gamma+e^{ik}),
\end{equation}
where $\Gamma$ is an $M \times  M$ Hermitian matrix defining the boundary Hamiltonian as
\[
H_{\rm boundary}=\Psi_0^\dag \;\Gamma \;
\Psi_0,
\]
with the notation defined in Sec.~\ref{section:Noninteracting_Conductance} right above Eq.~(\ref{Hamiltonian}). At half-filling, we have $k_F=\dfrac{\pi}{2}$ and the expression for the scattering matrix at the Fermi level reduces to 
\begin{equation}\label{eq:S_k_F}
    S_F=-\dfrac{\Gamma+i}{\Gamma-i}
\;.
\end{equation}
Note that it is important that the correct boundary condition is implemented at the Fermi level. Neither the conductance nor the behavior of the static $\langle J_R^\alpha(x) J_L^\beta(x) \rangle$ correlation functions are strongly affected by what happens deep inside the Fermi sea. It is evident from the form of the scattering matrix in Eq.~(\ref{eq:S_k_F}) that we need to change $\Gamma\rightarrow -\Gamma$ to invert the scattering matrix. In other words, if we have 
\begin{equation}\label{eq:H_L_H_R}
H_0=\Psi^\dagger_0 \Gamma_0 \Psi_0, \qquad H_\ell= \Psi^\dagger_\ell \Gamma_\ell \Psi_\ell,
\end{equation}
and we require $S_R(k_F)=S_0^{\dagger}(k_F)$ at the half-filling, we must have
\[
\Gamma_\ell=-\Gamma_0.
\]
This is one example where we in fact engineered the Hamiltonian $H_\ell$ to implement the boundary condition I. It is very helpful to find symmetry transformations that automatically reproduce $H_\ell$ from $H_0$. We argue that the transformation $\Gamma\rightarrow -\Gamma$ for the quadratic Hamiltonian $H_0$ is a combination of time-reversal and particle-hole transformations.

 Let us call these transformations $K$ and $C$, respectively. Since here we are considering spinless fermions, the time-reversal transformation is simply equal to the complex conjugation. Note that generically the time-reversal operator is anti-unitary and can be written as the product of the complex conjugation and a unitary operator. The particle-hole transformation simply switches the role of creation and annihilation operators. We can then write
 \begin{equation}
K(i)=-i, \qquad C(c)=c^\dagger, \qquad C(c^\dagger)=c.
 \end{equation} 

Acting with the particle-hole transformation on the left boundary Hamiltonian of the form Eq.~(\ref{eq:H_L_H_R}) yields 
\begin{eqnarray*}C(H_0)&=&C\left(\Psi^\dagger_0 \Gamma_0 \Psi_0\right)=\sum_{\alpha,\beta}^{M}C\left( \Gamma_0^{\alpha \beta}c^\dagger_{\alpha,0}c_{\beta, 0}\right)\\
&=&-\sum_{\alpha, \beta} \Gamma_0^{\alpha \beta}c^\dagger_{\alpha,0}c_{\beta,0}+{\rm const.}
\end{eqnarray*}
Neglecting an unimportant constant, we can write the above transformation in a more compact form as 
\[C(H_L)=-\Psi^\dagger_0 \Gamma^{\rm T}_0 \Psi_0=-\Psi^\dagger_0 \Gamma^{\rm *}_0 \Psi_0,\]
where we have used the Hermiticity of $\Gamma_0$ in the second step. We can then write
\begin{equation}\label{eq:symm_xform}
K(C(H_0))=-H_0=H_\ell.
\end{equation}

So far we have verified that for a simple junction described by the boundary Hamiltonian Eq.~(\ref{eq:H_L_H_R}), acting with $K$ and $C$ transformation on the boundary Hamiltonian inverts the Fermi-level scattering matrix at half-filling. We conjecture here that this is a more generic statement, and acting with these transformations on the boundary Hamiltonian gives a microscopic boundary Hamiltonian that correctly implements the same boundary conditions at both endpoints.
To support this conjecture, we test it on a more complex junction constructed by adding one more column of sites to the system. This calculation is presented in Appendix~\ref{App_A}.

The method developed in this paper can also be extended to study systems with spin-$1\over 2$ electrons. If there is no electron-electron interaction in the bulk of the wires, we can visualize the spin-up and down sectors as two distinct wires of spinless fermions. This construction can be used to engineer the Hamiltonian of the mirror-image junction from the scattering approach. In the presence of interactions in the bulk, the Hamiltonian of each wire can be written as a sum of a charge-sector Luttinger liquid with velocity $v_c$ and Luttinger parameter $g_c$ and a spin-sector Luttinger liquid with velocity $v_s$ and Luttinger parameter $g_s$. The conductance is then related to the the chiral current-current correlation function $\langle J_{cR}^\alpha(x) J_{cL}^\beta(x) \rangle$ in the charge sector. To measure this correlation function, we need to calculate $v_c$ in terms of the microscopic parameters such as the strength of the on-site Hubbard interaction $Un_\uparrow n_\downarrow$, where $\uparrow$ ($\downarrow$) indicates spin up (down).~\cite{Schulz90} This is a bulk property and can be calculated numerically or via the Bethe ansatz. The correlation function $\langle J_{cR}^\alpha(x) J_{cL}^\beta(x) \rangle$ can then be computed numerically using the following relations:
\[
J_\uparrow^\alpha+J_\downarrow^\alpha=v_c(J_{cR}^\alpha-J_{cL}^\alpha), \quad  N_\uparrow^\alpha+N_\downarrow^\alpha=J_{cR}^\alpha+J_{cL}^\alpha,
\]
which are analogous to Eq.~(\ref{eq:J_and_ N}). These relations lead to a similar equation to Eq.~(\ref{eq:measure}).

One final comment is in order regarding the efficient computation of the required correlation functions with DMRG. When interactions are confined to the junction itself, i.e., in the absence of interactions in the bulk of the wires, it is in some cases possible to perform a canonical transformation, which decouples some of the degrees of freedom in the system as recently shown in Ref.~\onlinecite{Feiguin2011}. Such transformations when possible can reduce the dimension of the local Hilbert space in DMRG calculations and lead to significant performance improvements.

\section{Numerical tests and benchmarks}
\label{section:Benchmarks}
\subsection{Noninteracting Y junction}
The first set of numerical tests and benchmarks we perform is with a noninteracting Y junction. The system we consider has three noninteracting fermionic chains with hopping amplitude of unity. At the junction, each chain is coupled to the other two chains with hopping amplitude $t$ and the loop formed at the junction is threaded with a magnetic flux $\Phi$. The gauge-invariant flux $\Phi$ can be modeled in many different ways in the tight-binding Hamiltonian such as, for example, the following boundary Hamiltonian:
\begin{equation}
 H_0=-\Psi^\dagger_0 \left(\begin{matrix}
   0& te^{i \Phi} & t  \\
  te^{-i \Phi}&0  & t \\
  t& t & 0
                          \end{matrix}
 \right)\Psi_0.
\end{equation}
Using Eq.~(\ref{S}) and the Landauer formula~(\ref{eq:Landauer}), we obtain the following exact expression for the conductance of this simple junction:
\begin{equation}\label{eq:exact_G}
  G_{12,21}=\frac{4 t^2 (1+t^2 \pm 2 t \sin \Phi)}{1+6 t^2 + 9 t^4 + 4 t^6 \cos^2 \Phi} \frac{e^2}{h}.
\end{equation}

\begin{figure}[htb]
\centering
\includegraphics[width=8 cm]{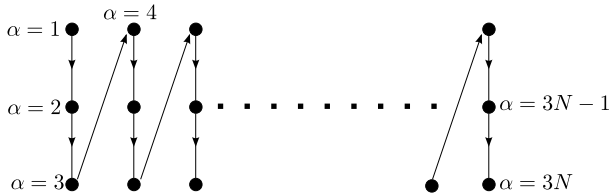}
\caption
{A labeling of system sites for exact diagonalization.}
\label{fig:label}
\end{figure}

We can now use exact diagonalization of the single-particle Hamiltonian, which can be written as an $3N\times 3N$ matrix (with $N$ the number of sites in each leg and $3N$ the total number of sites in the system), to obtain the $\langle J^\alpha(x) J^\beta(x)\rangle$ and $\langle N^\alpha(x) J^\beta(x)\rangle$ ground-state correlation functions. Notice that since the total Hamiltonian is quadratic, Wick's theorem [Eq.~(\ref{eq:wicks_theorem})] holds. Let us label all the $3N$ sites in the multi-wire system in some order as, for example, in Fig.~\ref{fig:label}, i.e., assign an annihilation operator $c_\alpha$, $\alpha=1\dots 3N$ to every site. All we need to calculate are then fermionic Green's functions of the form
\begin{equation}\label{eq:exact_C}
C(\alpha,\beta)=\langle c_\alpha^\dagger c_\beta \rangle.
\end{equation}

Diagonalizing the $3N \times 3N$ single-particle Hamiltonian gives single-particle wave functions $|\Psi_\epsilon\rangle=\sum_\alpha \phi^\epsilon_\alpha |\alpha\rangle$ where $|\alpha\rangle$ is a basis state of the single-particle Hilbert space with site $\alpha$ occupied and all other sites empty and $\epsilon$ represents the energy of the single-particle level. These wave functions are obtained directly from exact diagonalization. At a given filling $\nu$, the many-body wave function is a Slater determinant of $3\nu N$ low-energy single-particle wave functions. The wave functions of the filled levels are then $\phi^\epsilon_\alpha$ for $\epsilon<\epsilon_F$ where $\epsilon_F$ is the Fermi energy. We can then write 
\begin{equation}\label{eq:exact_G2}
C(\alpha,\beta)=\sum_{ \epsilon<\epsilon_F}\phi^{*\epsilon}_\alpha\phi^\epsilon_\beta.
\end{equation}

\begin{figure}[htb]
\centering
\includegraphics[width=8 cm]{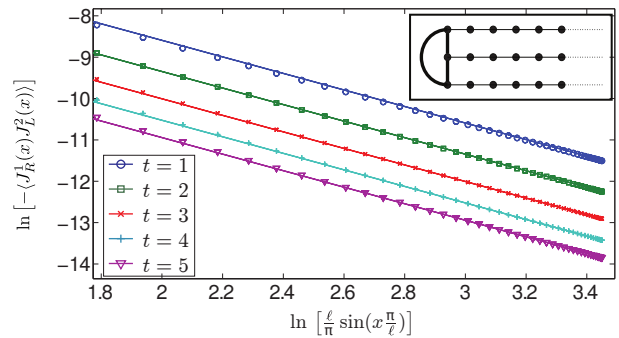}
\caption
{(Color online) The inset shows the noninteracting Y junction used for the numerical test of the method. The thick bonds at the junction have a different hopping amplitude $t$ than the thin bonds in the bulk of the wires, the hopping amplitude of which is $1$. The static correlation function is for a finite system of length $\ell=99$ ($100$ sites in each wire) for $\Phi=0$ and five different values of the hopping amplitude $t$. The lines drawn through the data points show the CFT prediction with the exact conductance Eq.~(\ref{eq:exact_G}). The strong agreement with the numerical data confirms that our method, i.e., extracting the conductance from fitting the correlators, yields accurate results.}
\label{fig:noninteracting_corr}
\end{figure}

We calculate, through exact diagonalization, the correlation function $\langle J^1_L(x) J^2_R(x)\rangle$ as explained above and plot their logarithms versus $\ln \left[ \frac{\ell}{\pi} \sin \left(x \frac{\pi}{\ell}\right)\right]$ for different values of the hopping $t$ and for two different fluxes $\Phi=0$ and $\Phi=\pi/2$. The results for $\Phi=0$ are shown in Fig.~\ref{fig:noninteracting_corr}.

\begin{figure}[htb]
\centering
\includegraphics[width=8 cm]{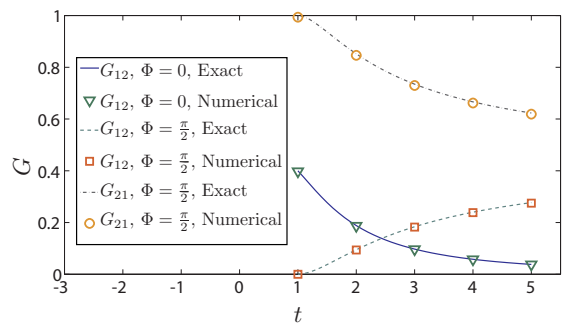}
\caption{(Color online) The conductance of the noninteracting Y junction calculated with our method versus the exact solution.}  
\label{fig:noninteracting_corr_G}
\end{figure}

As expected from the analysis of previous sections, far away from the junction, these plots are lines of slope $-2$. The conductance can be extracted from fitting the data in Fig.~\ref{fig:noninteracting_corr} and, as shown in Fig.~\ref{fig:noninteracting_corr_G}, it agrees with the exact conductance Eq.~(\ref{eq:exact_G}) with very good accuracy.

\subsection{Junction of two interacting quantum wires}

The next benchmark we perform is for two interacting wires with an impurity. As shown in Ref.~\onlinecite{Kane92a}, this system exhibits universal behavior. For attractive interactions, any impurity should heal at the RG fixed point, leading to a conductance of $g \frac{e^2}{h}$,  while for repulsive interactions, any impurity should effectively cut the chain resulting in zero conductance.

First, we show benchmarks for the attractive case. Here, of course, we are dealing with a strongly correlated system and exact diagonalization of the single-particle Hamiltonian can not be used. Due to the exponentially large dimension of the Hilbert space, we cannot exactly diagonalize the many-body Hamiltonian either. The DMRG method is an ideal tool for performing these calculations, and the results presented in this section are obtained by this method.

\begin{figure}[htb]
\centering
\includegraphics[width=8 cm]{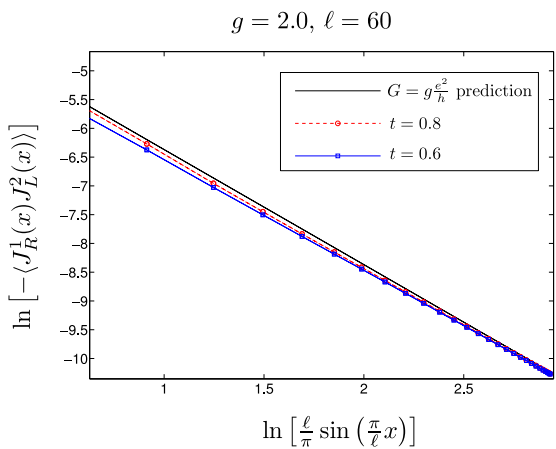}
\caption
{(Color online) The correlation functions are plotted for two values of hopping at the impurity $t=0.8$ and $t=0.6$ for the Luttinger parameter $g=2.0$. The strong agreement with the CFT prediction of $G=g {e^2 \over h}$ supports the correctness of our method.\newline
}
\label{fig:two_wire_g2}
\end{figure}

\begin{figure}[htb]
\centering
\includegraphics[width=8 cm]{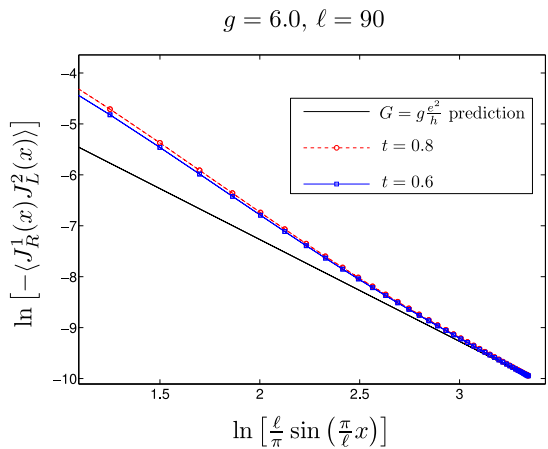}
\caption
{(Color online) The correlation functions are plotted for two values of hopping at the impurity $t=0.8$ and $t=0.6$ for the Luttinger parameter $g=6.0$. Here a larger healing length of $\ell=90$ is needed to observe the universal behavior and verify the CFT prediction.  }
\label{fig:two_wire_g6}
\end{figure}

The results of these calculations are shown in Figs.~\ref{fig:two_wire_g2} and \ref{fig:two_wire_g6} for the Luttinger parameter $g=2.0$ and $g=6.0$, respectively. For the larger Luttinger parameter $g=6.0$, we are very close to the phase separation point $V=2$ and we need a larger healing length. The universal behavior is evident from the fact that the value of the hopping amplitude at the junction has no effect on the asymptotic form of the correlation functions.

Let us now consider the case of repulsive interactions. This is a special case because the theoretical prediction for the conductance is $G_{12}=0$ so we do not expect to be able to fit the correlation function to the CFT form. The behavior of the correlation function, however, sheds light on the nature of the problem. As noted earlier, the CFT prediction for the universal form of the $\langle J_R^1 J_L^2 \rangle$ is an asymptotic form, and we expect subleading corrections coming from irrelevant boundary operators.

\begin{figure}[htb]
\centering
\includegraphics[width = 8cm]{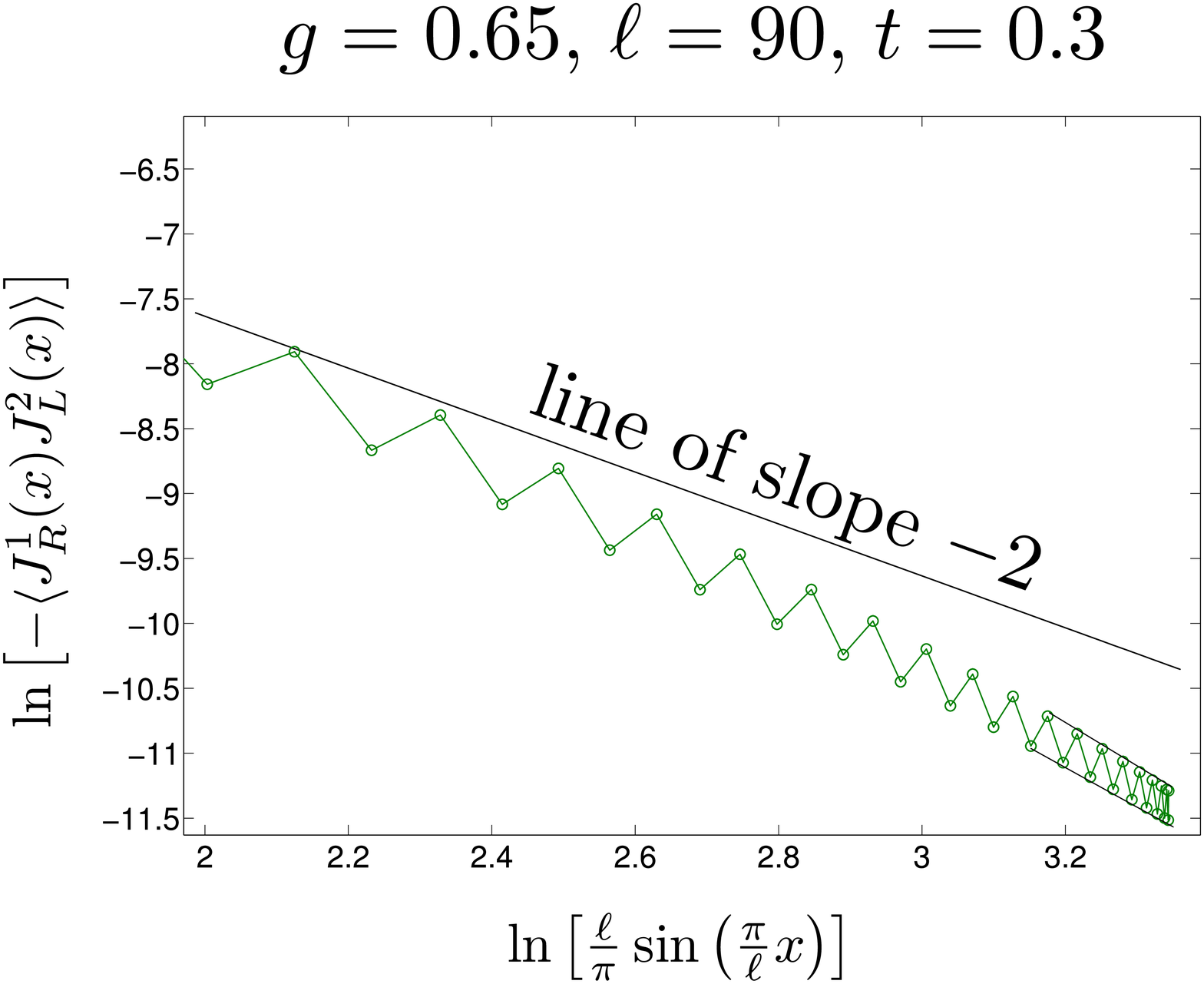}
\caption
{(Color online) The correlation functions are plotted for hopping at the impurity $t=0.3$ and the Luttinger parameter $g=0.65$. Since the universal conductance vanishes, we only observe the nonuniversal subleading corrections.}
\label{fig:repulsive}
\end{figure}

In the repulsive case, since the conductance is zero, $\langle J_R^1 J_L^2 \rangle$ will decay faster than $\left[\sin (x {\pi \over \ell}) \right]^{-2}$. As seen in Fig.~\ref{fig:repulsive} for $g=0.65 <1$, where the coefficient in front of the leading term is zero, we only observe these subleading corrections. As a function of $\ln \left[\sin (x {\pi \over \ell}) \right]$, $\ln \langle J_R^1 J_L^2 \rangle$ decays with a larger negative slope than $-2$ and in addition exhibits a $2 k_F$ oscillatory behavior.

The studies of this section verify that the method developed in this paper correctly reproduces established results in the presence or absence of electron-electron interactions. In the next section we apply the method to an open problem for three interacting wires.

\section{Universal conductance in an interacting Y junction of quantum wires}
\label{section:Y_Junction}

In this section, we employ the method developed in this paper to study a Y junction of interacting quantum wires with spinless electrons, each connected to a distinct site on a loop at the junction as seen in Fig.~\ref{fig:Yjunction}. The tunneling amplitude between the wires is $t$ and the flux through the loop is $\phi$. This system was analytically studied in Refs.~\onlinecite{Chamon03} and \onlinecite{Oshikawa06} where several nontrivial fixed points were identified. A similar analysis for a Y junction with spinful electrons is given in Ref.~\onlinecite{Hou08}.

\begin{figure}[htb]
\centering
\includegraphics[width=6.5 cm]{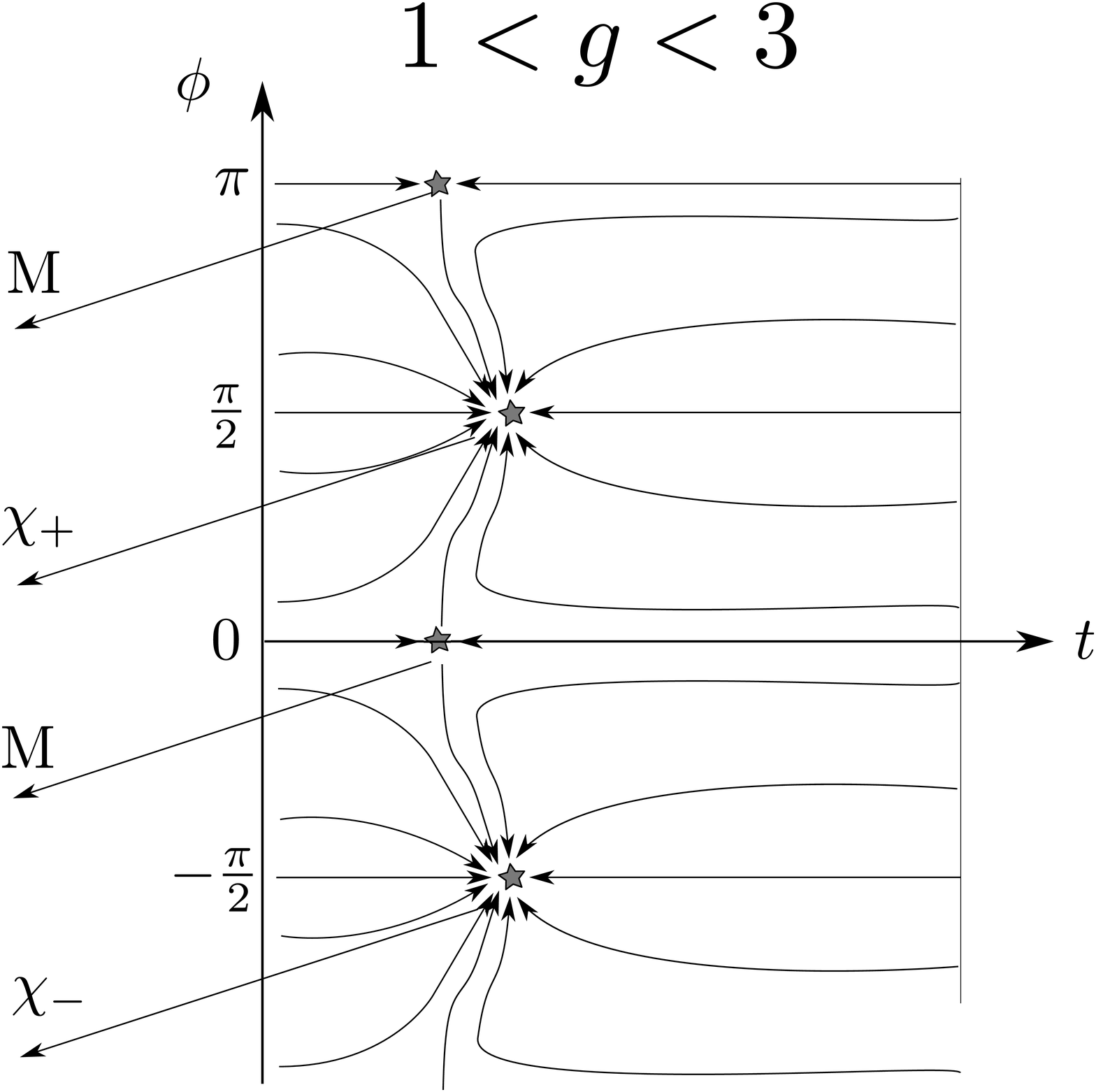}
\caption
{The conjectured RG flow diagram for the Y junction of Fig.~\ref{fig:Yjunction}, given in Ref.~\onlinecite{Oshikawa06}, for the Luttinger parameter in the range $1<g<3$.}
\label{fig:flow_diagram}
\end{figure}

In this work, we focus on the Luttinger parameter range $1<g<3$. The RG flow diagram for this range, which was conjectured in Ref.~\onlinecite{Oshikawa06}, is shown in the right-hand side of Fig.~\ref{fig:flow_diagram}. We have two types of fixed points: the chiral fixed points $\chi_+$ and $\chi_-$ in the presence of a time-reversal-symmetry-breaking flux and the M fixed point for unbroken time-reversal symmetry.

\subsection{Chiral fixed point}

\begin{figure}[htb]
\centering
\includegraphics[width= 8 cm]{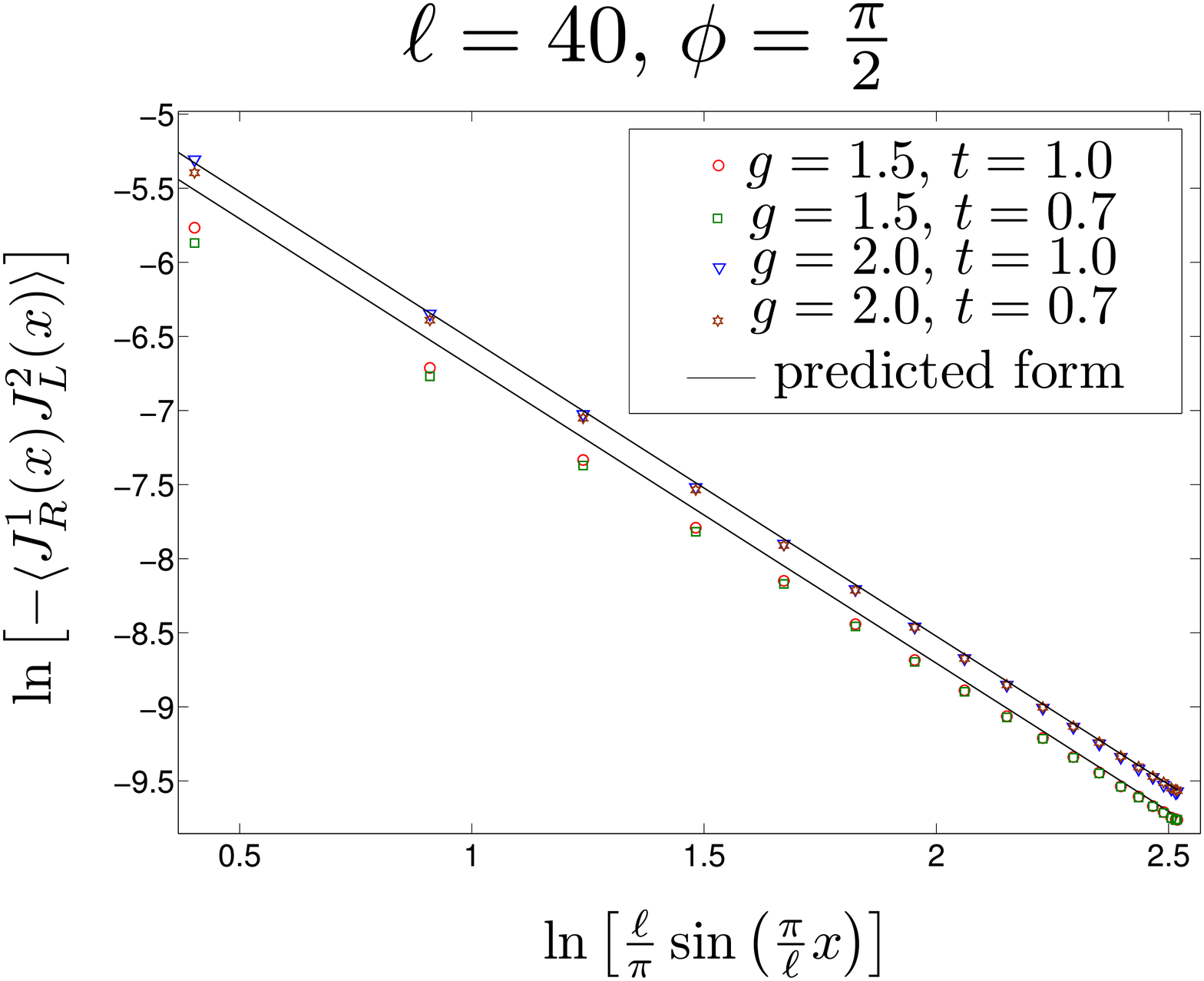}
\caption
{(Color online) A plot of the $\ln \langle J^1_R(x) J^2_L(x) \rangle$ for $g=1.5$ and $g=2.0$ and two different hopping amplitudes. We observe the universality of the conductance and good agreement with the theoretical predictions.}
\label{fig:G12L40}
\end{figure}

For the chiral fixed point, the time-reversal symmetry is broken by a magnetic flux and therefore $G_{12} \neq G_{21}$. It was predicted in Ref.~\onlinecite{Oshikawa06} that these conductances actually have a different sign and depend on the Luttinger parameter as follows:
\begin{eqnarray}\label{eq:G12}
G_{12}&=&-2{g \over(3+g^2)} (g+1){e^2\over h},\\
G_{21}&=&2{g \over(3+g^2)} (g-1){e^2\over h}.
\end{eqnarray}
This conductance is universal, i.e., it is independent of the hopping amplitude $t$ at the junction for nonvanishing $t$. It is worthwhile to emphasize that the emergence of universal conductances strictly relies on the presence of interactions, i.e., by taking the noninteracting limit $g\to 1$, we shall not expect to recover correct results for noninteracting cases.~\cite{note1} In the RG language, the noninteracting point is a marginal fixed point and each value of hopping amplitude $t$ gives a unique conductance as in Eq.~(\ref{eq:exact_G}). Because we are studying junctions with $Z_3$ symmetry, the conductances obey the relations, $G_{12}=G_{23}=G_{31}$ and $G_{21}=G_{32}=G_{13}$. In the time-reversal symmetric case (the M fixed point), the conductances in addition follow $G_{ij}=G_{ji}$. To verify these predictions, we numerically calculated the correlation function, $\langle J^1_R(x) J^2_L(x) \rangle$, for two values of $t$ and two values of $g$ in a system size of $\ell=40$. As shown in Fig.~\ref{fig:G12L40}, the numerical results strongly agree with the prediction for $G_{12}$. The consistency serves both as the first numerical verification of the prediction Eq.~(\ref{eq:G12}) as well as a highly nontrivial benchmark for the correctness of our method.

\begin{figure}[htb]
\centering
\includegraphics[width= 8cm]{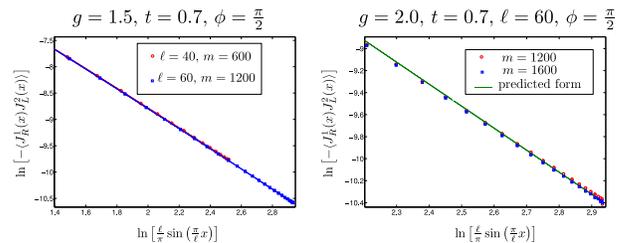}
\caption
{(Color online) In the left panel, we check the convergence of the correlation function with increasing system size for $g=1.5$ and $t=0.7$. As seen, the correlation functions collapse for $\ell=40$ and $\ell=60$. In the right panel, we check the DMRG convergence as the truncation error is reduced by increasing $m$. The data are for $g=2.0$ and $t=0.7$ in a system size of $\ell=60$.}
\label{fig:convergence}
\end{figure}

It is quite remarkable that we can verify the $G_{12}$ conductance in a system as small as $\ell=40$. Note that due to the asymptotic form of the relation between the universal conductance and the computed correlation functions, the results are reliable once there is convergence in the system size. Also, one needs to check for the convergence of the DMRG method by increasing the number of states used in the calculations to make sure the DMRG results are in fact quasi-exact. We show some of these convergence tests in Fig.~\ref{fig:convergence}.

In the last part of this section, we present the numerical verification of the $G_{21}$ conductance. It turns out that for $G_{21}$, a longer healing length is needed and we are able to see the agreement with the theoretical prediction in a system of $\ell=60$. It is remarkable that the correlation function here actually changes sign and the universal long-distance behavior, corresponding to a positive conductance, has opposite sign to the nonuniversal values of the correlation function near the junction.

\begin{figure}[htb]
\centering
\includegraphics[width= 8cm]{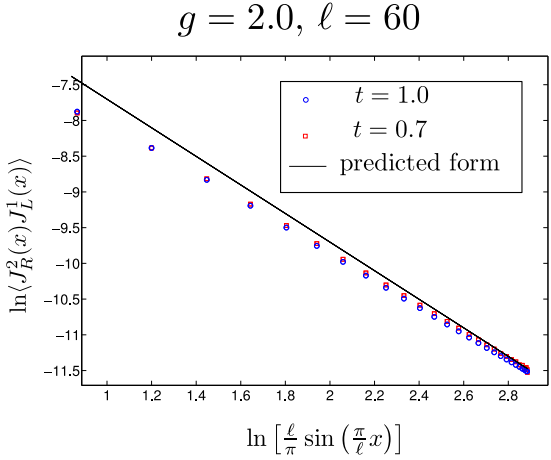}
\caption
{(Color online) The universal $G_{21}$ conductance for $g=2.0$ and two different values of the hopping amplitude $t=1.0$ and $t=0.7$. Notice the sign of the argument of the logarithmic function labeling the vertical axis. Here, $G_{12}$ is positive and the correlation function $\langle J_R^2(x)J_L^1(x) \rangle$ actually changes sign at a certain distance from the boundary.}
\label{fig:G21}
\end{figure}

Due to the sign change of the correlation function, here we plot the logarithm of minus the correlation function in a region far away from the junction (where the argument of the logarithm is positive). The results are shown in Fig.~\ref{fig:G21} and, once again, there is good agreement between the data and the theoretical prediction.

\subsection{M fixed point}

Let us now turn to the conductance of the time-reversal symmetric M fixed point for which theoretical predictions do not exist. The benchmarks in the previous section and the study of the chiral fixed point for three interacting wires increases our confidence in the correctness of the method developed here. We are now ready to apply the method to a thus-far unsolved quantum impurity problem.

\begin{figure}[htb]
\centering
\includegraphics[width = 8cm]{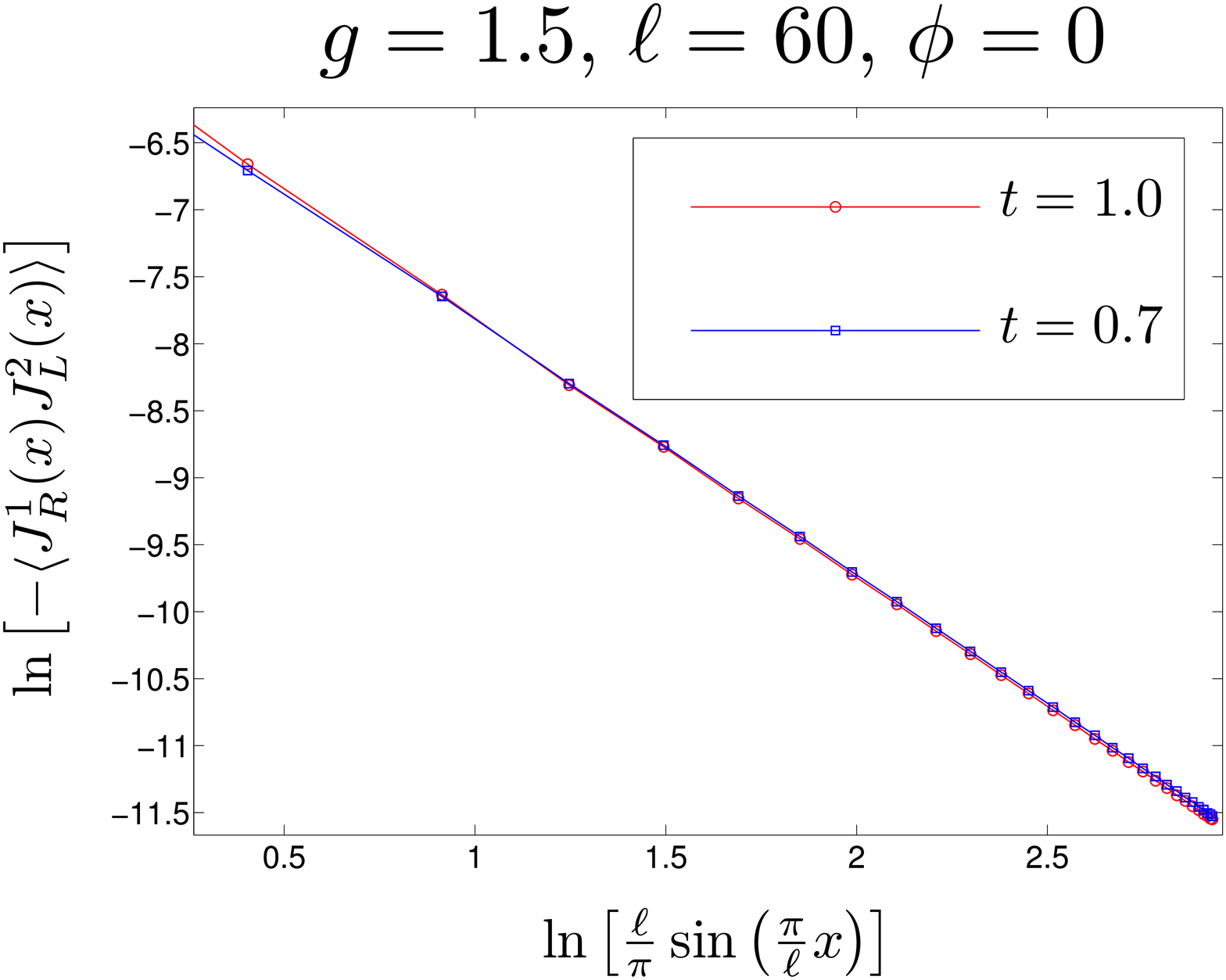}
\caption
{(Color online) The universal $G_{12}=G_{21}$ conductance for $g=1.5$ and two different values of the hopping amplitude $t=1.0$ and $t=0.7$.}
\label{fig:g15L60}
\end{figure}

\begin{figure}[htb]
\centering
\includegraphics[width=8cm]{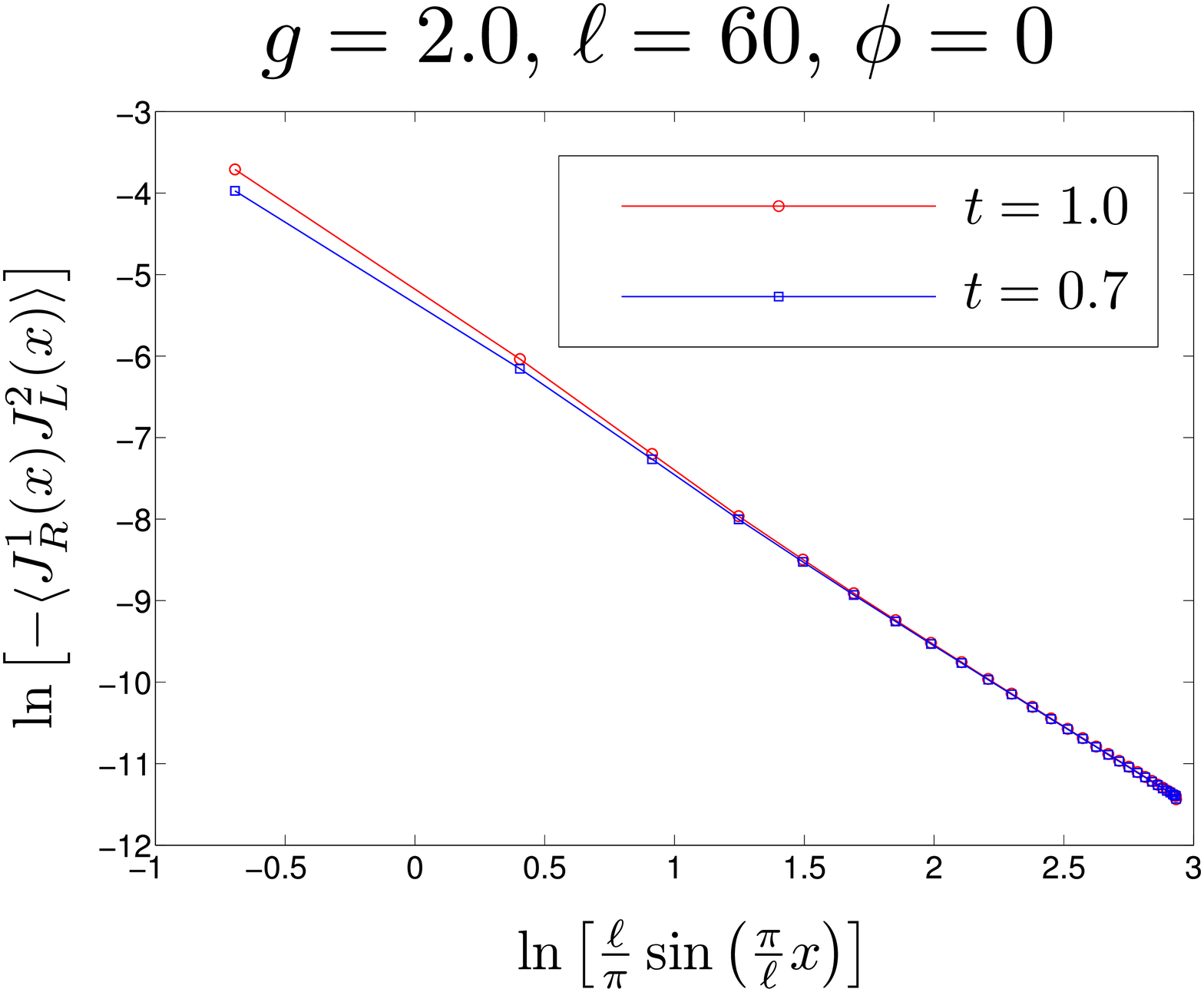}
\caption
{(Color online) The universal $G_{12}=G_{21}$ conductance for $g=2.0$ and two different values of the hopping amplitude $t=1.0$ and $t=0.7$.}
\label{fig:g20L60}
\end{figure}

\begin{figure}[htb]
\centering
\includegraphics[width=8cm]{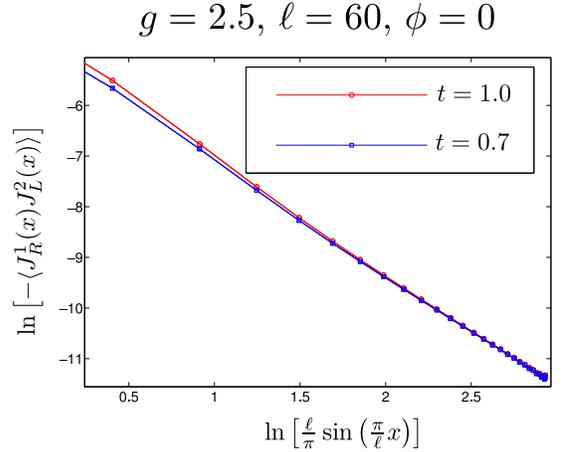}
\caption
{(Color online) The universal $G_{12}=G_{21}$ conductance for $g=2.5$ and two different values of the hopping amplitude $t=1.0$ and $t=0.7$.}
\label{fig:g25L60}
\end{figure}

We perform the calculations of the M fixed-point conductance for three values of the Luttinger parameter $g$ and two values of the hopping amplitude $t$. The independence from the hopping amplitude can be seen at $\ell=60$. The results are shown in Figs.~\ref{fig:g15L60},~\ref{fig:g20L60}, and~\ref{fig:g25L60}, respectively for $g=1.5, 2.0, 2.5$. 

\begin{figure}[htb]
\centering
\includegraphics[width=8cm]{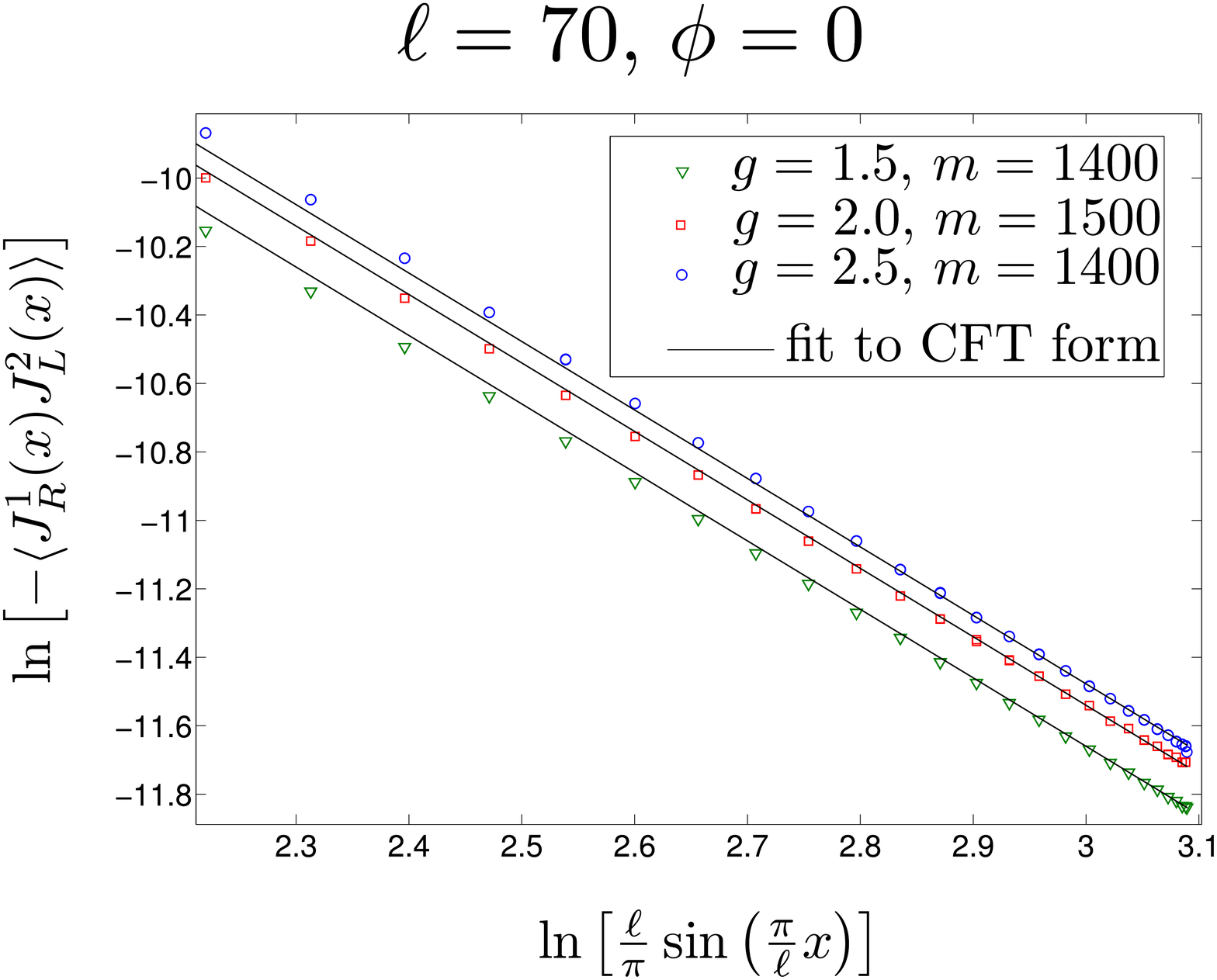}
\caption
{(Color online) Fitting the correlation function to the CFT prediction to extract the conductance. The data in this figure are for $t=0.7$ and $\ell=70$, with three different values of $g$.}
\label{fig:L70}
\end{figure}

To obtain the conductance, we fit the data for a slightly larger system of $\ell=70$. The three correlation functions are plotted in Fig.~\ref{fig:L70} in addition to the best fit in the asymptotic region for each value of $g$. We obtain the following conductances for the M fixed point:
\begin{eqnarray*}
G_{12}(g=1.5)&=&G_{21}(g=1.5)\approx -0.55 \:{e^2 \over h},\\
G_{12}(g=2.0)&=&G_{21}(g=2.0)\approx -0.62 \:{e^2 \over h},\\
G_{12}(g=2.5)&=&G_{21}(g=2.5)\approx -0.665 \:{e^2 \over h}.
\end{eqnarray*}

Is there a pattern to these values of the conductance? It is well known that when a Luttinger liquid with attractive interactions ($g>1$) is coupled to Fermi-liquid leads, the $g e^2/ h$ conductance renormalizes to $e^2/ h$.~\cite{Maslov95,Safi05} One way to think about this phenomenon is to assume that we have an effective contact resistance at the interface between each interacting wire and the noninteracting leads. If this contact resistance is $1/g_c$, we must have
\[
1/g+2/g_c=1 \quad \Rightarrow \quad g_c=2/(1-g^{-1}).
\]
For the multi-wire conductance tensor, we can consider the following transformation, which, if the effective contact resistance is the correct mechanism, renormalizes the conductance in the presence of noninteracting leads. Here, we think of the transformation
 \[
\bar{G}=\left(\openone+G/g_c\right)^{-1}G
\]as a mathematical mapping of the conductance tensor,
which can be inverted as 
\[
G=\bar{G} \left(\openone-\bar{G}/g_c\right)^{-1}.
\]
 
 It turns out that for all the other fixed points of the Y junction as well as in the two-wire case, $\bar{G}$ is independent of the Luttinger parameter.~\cite{Oshikawa06} We conjecture here that this is also true for the M fixed point. We can check explicitly that for the three values of $g$ studied above,
\[
\bar{G}\approx\left(\begin{matrix}
2\:\gamma & -\gamma  & -\gamma \\ 
-\gamma & 2\:\gamma & -\gamma \\ 
-\gamma & -\gamma  & 2\:\gamma 
\end{matrix}\right){e^2 \over h}
\] 
with $\gamma\approx 0.42$. This conjecture leads to a general form for the conductance of the M fixed point as a function of the Luttinger parameter and one dimensionless number $\gamma\approx 0.42$ as
\[
G_{12}(g)=G_{21}(g)=-\frac{2 g \gamma}{2g+3\gamma-3g\gamma}{e^2 \over h}.
\]

\begin{figure}[htb]
\centering
\includegraphics[width=8cm]{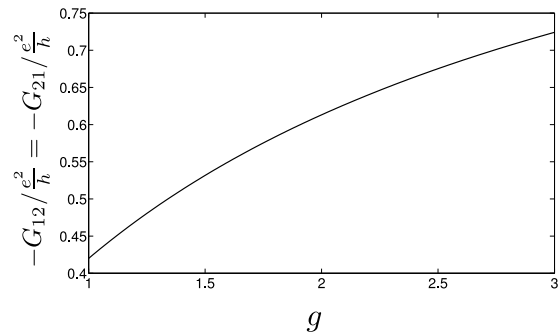}
\caption
{The dependence of the off-diagonal conductance of the M fixed point on the Luttinger parameter $g$.}
\label{fig:M_G}
\end{figure}

As stated above, this conjecture works to a good approximation for the three values of $g$ studied above numerically. The functional form of the dependence of the conductance of the M fixed point on the Luttinger parameter is plotted in Fig.~\ref{fig:M_G}. This simple plot, which had remained elusive since the prediction of the existence of the M fixed point in~\cite{Chamon03}, is the main result of this section and an important achievement for the method developed in this paper.

Notice that the measured value of $\gamma$ differs from $4/9\approx0.444$ by about $4.8\%$. There are several sources of error in our evaluation of $\gamma$. The DMRG truncation error is very small in this case and the calculated correlation functions are quasi-exact. The error from fitting the data is around $2\%$ and there are errors from finite-size corrections that with our system sizes are of the order a few percent. It seems plausible that the renormalized conductance of the M fixed point, when attached to noninteracting leads, is equal to $G_{12}=-{4 \over 9} e^2/h$ as indicated by the results of Ref.~\onlinecite{Barnabe05b} which were obtained by an approximate functional renormalization group scheme. This strongly suggests a simple understanding of the M fixed point conductance because ${4 \over 9} e^2/h$ is the largest conductance $|G_{12}|$ we can have, with time reversal symmetry, for three noninteracting quantum wires connected by a unitary scattering matrix. The $g$-dependence of the M fixed point conductance is such that when attached to noninteracting leads the system acts as an effective scattering matrix with the largest possible conductance $G_{12}$. 
 
\section{Conclusions and outlook}
\label{section:conclusions}
In summary, we have developed a method in this paper that allows the computation of the conductance tensors of rather arbitrary quantum junctions with standard numerical techniques such as time-independent DMRG. We presented specific studies of an interacting Y junction and resolved an outstanding open question regarding the conductance of the M fixed point.

Our generic method has only one practical limitation. Namely, the size $\ell$ of this system must be large enough for the universal scaling behavior to emerge. Two factors affect this minimum length. First, an intrinsic healing length required to observe the continuum long-distance Luttinger-liquid behavior. For large attractive interactions (large $g$), this intrinsic healing length is large. Second, the scaling dimension of the irrelevant boundary operators. If these operators are only slightly irrelevant, we need to move away a larger distance from the boundary to observe the fixed-point behavior. The $D_p$ fixed point of the Y junction, for example, which occurs for $3<g<9$, presents these difficulties. The scaling dimension of the leading irrelevant boundary operator increases with $g$, but making $g$ larger brings the Luttinger liquid close to the phase separation point and increases the intrinsic healing length. We have attempted to verify the theoretical prediction for the $D_p$ fixed point but have not been able to reach the required length with our computational resources. As demonstrated in this paper, however, this practical limitation is by no means a generic problem.

The method developed in this paper opens up the possibility of applying the powerful technology of DMRG and matrix product states, which in the last decades have been successfully applied to quasi-1D systems such as ladders, to transport calculations in the presence of electron-electron interactions. For decades, quantum transport was formulated for noninteracting electrons, and interaction effects were treated by mean-field and approximate methods. The results of this paper make it possible to utilize the quasi-exact numerical methods such as time-independent DMRG to study the transport properties of strongly correlated systems.

Many interesting questions can now be answered with standard computations. A long-standing open question is the conductance of a chain of interacting spinful fermions at the nontrivial Kane and Fisher fixed point.~\cite{Kane92,Furusaki93,Wong94} The method also makes it possible to study fixed points of more than three quantum wires. For two wires of spinless fermions, there are only two interacting fixed points, while the Y junction presents a much richer flow diagram with multiple fixed points.  What are the possible universal conductances with four, five wires, etc.? Are there hierarchies or other mathematical structures?

Another direction of interest is to identify the universality class that particular junctions, which are experimentally accessible, fall into. Transport measurements are done on a multitude of organic molecules such as fullerenes, for example. Effective tight-binding models can be derived for such molecular structures via first-principles calculations. Our method combined with DMRG computations is then helpful for extracting the universal behavior of such molecular junctions.

In summary, the technology developed in this paper provides a systematic way to study the effects of strong electron-electron interactions in the transport properties of quantum impurity problems and molecular electronic devices. In this framework, and with the help of efficient DMRG computations, strong correlations can be studied in a quasi-exact manner without resorting to mean-field or perturbative treatments.    

\acknowledgements
{We are grateful to A. Polkovnikov and A. Sandvik for
helpful comments and discussions. This work was supported in part by the DOE Grant
No. DE-FG02-06ER46316 (CC, CH, and AR), the Dutch Science Foundation
NWO/FOM (CH), NSF DMR-0955707 (AF), KAKENHI No. 50262043 (MO),
NSERC (IA) and CIfAR (IA).}

\appendix
\section{Symmetry transformation for a two-column noninteracting junction}
\label{App_A}
\begin{figure}
 \centering
 \includegraphics[width= 4 cm]{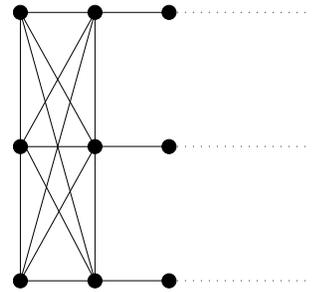}
 \caption{A junction with two columns of fermionic sites at the boundary and a general quadratic boundary Hamiltonian as in Eq.~(\ref{eq:two_column}).}
 \label{fig:jun}
\end{figure}
In this appendix, we test the $H_R=K(C(H_L))$ conjecture of Eq.~(\ref{eq:symm_xform}) for a two-column noninteracting junction. 
The boundary Hamiltonian for the junction, which is shown in Fig.~\ref{fig:jun}, is as follows: 
\begin{equation}\label{eq:two_column}
H_L=\Psi^\dagger_0 \Gamma_0 \Psi_0+\Psi^\dagger_{-1} \Gamma_{-1} \Psi_{-1}+\Psi^\dagger_{-1}T \Psi_{0}+\Psi^\dagger_{0}T^\dagger \Psi_{-1},
\end{equation}
where the two $\Gamma$ matrices are Hermitian and $T$ is an arbitrary $M\times M$ matrix. Let us first calculate the $S$ matrix. We have
\begin{eqnarray}
&&\epsilon\psi_{-1}=\Gamma_{-1}\psi_{-1}+T \psi_{0},\label{eq:equation_of_motion1}\\
  &&\epsilon\psi_{0}=\Gamma_{0}\psi_{0}+T^\dagger \psi_{-1}-\psi_{1}.\label{eq:equation_of_motion2}
\end{eqnarray}
We work at half-filling where $\epsilon=0$. From Eq.~(\ref{eq:equation_of_motion1}), we then obtain
\[\psi_{-1}=-\left(\Gamma_{-1}\right)^{-1}T \psi_0.\]
Plugging this into Eq.~(\ref{eq:equation_of_motion2}) and replacing the wave functions $\psi$ with scattering plane waves gives
\begin{eqnarray*}
\Gamma_0(A_{\rm out}+A_{\rm in})&-&T^\dagger \left(\Gamma_{-1}\right)^{-1}T(A_{\rm out}+A_{\rm in})\\
&-&i(A_{\rm out}-A_{\rm in})=0
\end{eqnarray*}
from which we get the following scattering matrix:
\begin{eqnarray*}
S&=&\\
& &\left[i-\Gamma_0+T^\dagger \left(\Gamma_{-1}\right)^{-1}T\right]^{-1}\left[i+\Gamma_0-T^\dagger \left(\Gamma_{-1}\right)^{-1}T\right].
\end{eqnarray*}

We have seen already that the transformation $KC$ takes $ \Gamma_0$ to $-\Gamma_0$. It similarly changes $ \Gamma_{-1}$ to $-\Gamma_{-1}$. In order for this transformation to work, we need to have $T^\dagger \left(\Gamma_{-1}\right)^{-1}T\rightarrow-T^\dagger \left(\Gamma_{-1}\right)^{-1}T$ under $KC$. We will show this by arguing that the matrix $T$ goes to $-T$. Let us begin by applying the particle-hole transformation to the $T$ term $\Psi^\dagger_{-1}T \Psi_{0}+\Psi^\dagger_{0}T^\dagger \Psi_{-1}$. We have
\begin{eqnarray*}
&&C(\Psi^\dagger_{-1}T \Psi_{0}+\Psi^\dagger_{0}T^\dagger \Psi_{-1})\\
&=&-\sum_{j,j'}\left(T_{jj'}c_{-1j}c^\dagger_{0j'}+T^\dagger_{jj'}c_{0j}c^\dagger_{-1j'}\right)\\
&=&-\Psi^\dagger_{-1}T^* \Psi_{0}-\Psi^\dagger_{0}T^{\rm T} \Psi_{-1}
\end{eqnarray*}
and, therefore, 
\begin{eqnarray*}
&&K\left(C(\Psi^\dagger_{-1}T \Psi_{0}+\Psi^\dagger_{0}T^\dagger \Psi_{-1})\right)\\
&=&\Psi^\dagger_{-1}(-T) \Psi_{0}+\Psi^\dagger_{0}(-T^\dagger )\Psi_{-1}.
\end{eqnarray*}
So, we find that indeed $T^\dagger \left(\Gamma_{-1}\right)^{-1}T\rightarrow-T^\dagger \left(\Gamma_{-1}\right)^{-1}T$.
\section{Relationship between $A^{\alpha\beta}_{\cal B}$ and the boundary state $|{\cal B}\rangle$}
\label{app:highest_weight}
In this appendix, we review the notion of the highest-weight states and derive Eq.~(\ref{eq:highest_weight}), which explicitly relates the coefficient $A^{\alpha \beta}_{\cal B}$ to the boundary state and the highest-weight states of some primary operators in the CFT. The derivation in this appendix is merely a review of the generic result obtained in Ref.~\onlinecite{Cardy_Lewellen91} and is presented for completeness.

In field theory, it is customary to quantize the theory for constant times. The Hamiltonian of the system is the generator of time translation. In CFT, however, because of scale invariance, it is convenient to quantize the theory on a fixed circle in the complex $z=\tau+i x$ plane. This is called radial quantization and the generator of scale transformations is called the dilatation operator $D$, which can be written as the integral of the radial component of the stress tensor. 

The operator-state correspondence in CFT is formulated in the radial quantization framework. There is a vacuum state $|0\rangle$ in radial quantization, and acting by the operator corresponding to a scaling field on the vacuum gives a state
\[
 | \phi \rangle=\hat{\phi}(0,0) |0 \rangle.
\]
It can be shown that these states are eigenstates of the dilatation operator $D.$~\cite{Cardy10} The state constructed in this manner for a primary field is called the highest-weight state of that field. 

Let us now turn to the derivation of Eq.~(\ref{eq:highest_weight}) following Ref.~\cite{Cardy_Lewellen91}. We would like to show that if $\langle O(x) \rangle=A^O_{\cal B} (2x )^{-X_O}$ for a scaling operator $O$ living on the upper-half plane, we have
\[
 A^O_{\cal B}=\frac{\langle O|{\cal B} \rangle}{\langle 0|{\cal B} \rangle},
\]
where $|{\cal B} \rangle$ is the boundary state on the real axis.

\begin{figure}[htb]
\centering
\includegraphics[width =5 cm]{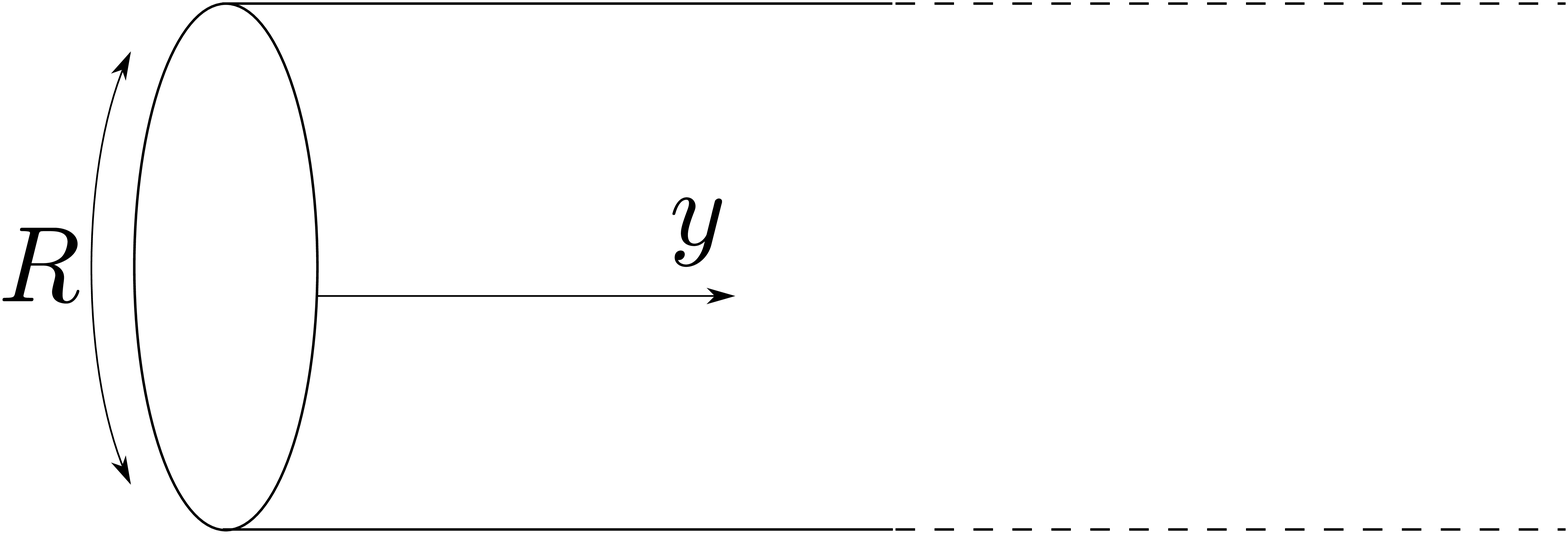}
\caption
{The expectation value of $O$ is calculated on the semi-infinite cylinder of radius $R$ in two different ways.}
\label{fig:semi-cyl}
\end{figure}

The key to this derivation is calculating the expectation value of $O$ on a semi-infinite cylinder of radius $R$ seen in Fig.~\ref{fig:semi-cyl} in two different ways. First, we use a conformal mapping from the upper half-plane to the semi-infinite cylinder. Consider $\tilde{w}=\frac{1-iz}{1+iz}$. It is easy to show that for $z$ on the real axis, $\tilde{w} \tilde{w}^*=1$ and the upper half-plane is mapped to the outside of a unit disk. It is then easy to see that the conformal mapping 
\[
 w(z)=\frac{R}{2\pi}\ln\left(\frac{1-iz}{1+iz} \right) 
\]
takes the upper-half plane to the semi-infinite cylinder.
 By applying this transformation to the expectation value of $O$ on the semi-infinite plane, we obtain
 \begin{equation}\label{eq:with_mapping}
 \langle O(y) \rangle=A^O_{\cal B}\left( \frac{2\pi}{R}\right)^{X_O}\frac{\exp(-2\pi X_O y/R)}{\left[ 1-\exp(-4\pi y/R)\right]^{X_O} }.
\end{equation}

Now a second way to calculate this quantity is directly on the semi-cylinder and by using the boundary state. If the transfer matrix in the $y$ direction (parallel to the cylinder axis) is $\exp (- {\cal H})$ for a Hamiltonian $\cal H$, we can write
\[
 \langle O(y) \rangle=\lim_{T\rightarrow\infty}\frac{\langle 0|e^{-(T-y){\cal H}}O(y)e^{-y{\cal H}}|{\cal B}\rangle}{\langle 0|e^{-T{\cal H}}|{\cal B}\rangle},
\]
where $|0\rangle$ is the ground state of $\cal H$.

We now insert the resolution of the identity between $O(y)$ and $e^{-y{\cal H}}$ in the above expression. Since $O$ is a primary operator only states that are not annihilated by $O$ enter the sum. The highest-weight state $O |0\rangle=|O \rangle$ has the smallest eigenvalue ~\cite{Cardy10} and, in the limit of large $y$, will be the leading term. We then obtain
\begin{equation}\label{eq:directly}
 \langle O(y) \rangle=\frac{\langle 0|O|O\rangle\langle O|{\cal B}\rangle}{\langle 0|{\cal B}\rangle} e^{-\varepsilon_O y},
\end{equation}
 where $\varepsilon_O=2\pi  X_O /R$.~\cite{Cardy_Lewellen91} We now argue that $\langle 0|O|O\rangle $ behaves as $(2 \pi /R)^{X_O}$. This can be seen from calculating $\langle 0|O(y_1)O(y_2)|0\rangle$ on a cylinder. By mapping the full complex plane to a cylinder of radius $R$, we obtain~\cite{Cardy86}
\[
 \langle 0|O(y_1)O(y_2)|0\rangle=\left[ \frac{R}{\pi} \sinh \frac{\pi}{R}(y_1-y_2) \right]^{-2 X_O}.
\]
 
In the limit of $(y_1-y_2)\rightarrow\infty$, the above correlation function reduces to 
\[
\langle 0|O(y_1)O(y_2)|0\rangle\sim\left( \frac{2\pi}{R}\right) ^{2 X_O}e^{-\frac{2\pi}{R} X_O (y_1-y_2)}.
\]  
Alternatively, we can write an expression for the above correlation function by inserting the resolution of the identity between $Q(y_1)$ and $O(y_2)$ and taking the limit of $(y_1-y_2)\rightarrow\infty$, which yields
\[
\langle 0|O(y_1)O(y_2)|0\rangle\sim |\langle 0|O|O\rangle|^2 e^{-\varepsilon_O (y_1-y_2)}.
\]
We then find by comparison that 
\begin{equation}\label{eq:matrix_element}
\langle 0|O|O\rangle \sim(2 \pi /R)^{X_O}.
\end{equation}
Comparing Eq.~(\ref{eq:with_mapping}) (in the limit of large $R$) and Eq.~(\ref{eq:directly}) and making use of the above expression Eq.~(\ref{eq:matrix_element}) gives the following final relationship between the coefficient $A^O_{\cal B}$ in terms of the boundary state:
\[
A^O_{\cal B}=\frac{\langle O|{\cal B}\rangle}{\langle 0|{\cal B}\rangle}.
\]
 
\bibliography{junction}

\begin{thebibliography}{67}
\bibitem[{not()}]{note1}
\bibinfo{note}{Indeed it is interesting to observe that the noninteracting
  limit, $g\to1$, of Eq.~\eqref{eq:G12} coincides with the noninteracting
  conductances in Eq.~(\ref{eq:exact_G}) with $t=1$ and $\Phi=\pi/2$.}
\end{thebibliography}

\end{document}